\documentclass[aps,pra,twocolumn,groupedaddress,showpacs]{revtex4}
\usepackage{graphicx}
\begin{document}

\def \tr{{\mbox{tr~}}}
\def \ra{{\rightarrow}}
\def \ua{{\uparrow}}
\def \da{{\downarrow}}
\def \l{\left}
\def \r{\right}
\def \half{{1\over 2}}
\def \etal{{\it {et al}}}
\def \cH{{\cal{H}}}
\def \cM{{\cal{M}}}
\def \cN{{\cal{N}}}
\def \cQ{{\cal Q}}
\def \cI{{\cal I}}
\def \cV{{\cal V}}
\def \cG{{\cal G}}
\def \cF{{\cal F}}
\def \cZ{{\cal Z}}
\def \bS{{\bf S}}
\def \bI{{\bf I}}
\def \bL{{\bf L}}
\def \bG{{\bf G}}
\def \bQ{{\bf Q}}
\def \bK{{\bf K}}
\def \bR{{\bf R}}
\def \br{{\bf r}}
\def \bu{{\bf u}}
\def \bq{{\bf q}}
\def \bk{{\bf k}}
\def \bz{{\bf z}}
\def \bx{{\bf x}}
\def \bpsi{{\bar{\psi}}}
\def \tJ{{\tilde{J}}}
\def \W{{\Omega}}
\def \e{{\epsilon}}
\def \lam{{\lambda}}
\def \L{{\Lambda}}
\def \a{{\alpha}}
\def \t{{\theta}}
\def \b{{\beta}}
\def \g{{\gamma}}
\def \D{{\Delta}}
\def \d{{\delta}}
\def \w{{\omega}}
\def \s{{\sigma}}
\def \f{{\varphi}}
\def \x{{\chi}}
\def \e{{\epsilon}}
\def \h{{\eta}}
\def \G{{\Gamma}}
\def \z{{\zeta}}
\def \hatt{{\hat{\t}}}
\def \hn{{\bar{n}}}
\def \vk{{\bf{k}}}
\def \vq{{\bf{q}}}
\def \gk{{\g_{\vk}}}
\def \nd{{^{\vphantom{\dagger}}}}
\def \yd{^\dagger}
\def \av#1{{\langle#1\rangle}}
\def \ket#1{{\,|\,#1\,\rangle\,}}
\def \bra#1{{\,\langle\,#1\,|\,}}
\def \braket#1#2{{\,\langle\,#1\,|\,#2\,\rangle\,}}

\newcommand{\bea}{\begin{eqnarray}}
\newcommand{\eea}{\end{eqnarray}}
\newcommand{\nn}{\nonumber}
\newcommand{\rav}{\rangle}
\newcommand{\lav}{\langle}
\newcommand{\bfr}{{\bf r}}
%\newcommand{\ua}{\uparrow}
%\newcommand{\da}{\downarrow}
%\preprint{}


\title{Phase diagrams of one-dimensional Bose-Fermi 
mixtures of ultra-cold atoms}



\author{L. Mathey$^1$ and D.-W. Wang$^2$}

%\thanks{}

\affiliation{${}^1$Physics Department, Harvard University, Cambridge, MA 02138 \\
${}^2$Department of Physics, National Tsing-Hua University, Hsinchu, Taiwan 300, Republic of China}


\date{\today}

\begin{abstract}
We study the quantum phase diagrams of Bose-Fermi mixtures 
of ultracold atoms confined to one dimension 
 in an optical lattice. 
For systems 
 with incommensurate densities,  
various quantum 
phases, e.g. charge/spin density waves, pairing, 
%the Wentzel-Bardeen
%instability, 
 phase separation, 
and the Wigner crystal,
are found to be dominant in different parameter regimes
within a bosonization approach.
The structure of the phase diagram leads us to propose that the
system is best understood
as a Luttinger liquid of polarons (i.e. atoms of one species
surrounded by screening clouds of the other species).
%, which is 
% consistent with a renormalization group analysis of a local impurity potential.
Special fillings, half-filling for fermions and unit filling for bosons,
 and the resulting gapped phases are also discussed, as well as
 %
% 
%Taking commensurabilities into account, we find that the system
% can develop different types of gapped phases.
%By applying a renormalization group analysis for a local impurity potential,
%we exclude the simple metallic phase in the region between spin/charge
%density waves and the fermionic pairing phases, showing that
%the entire phase diagram is best understood
%as a Luttinger liquid of polarons (i.e. atoms of one species
%surrounded by screening clouds of other species).
%We also discuss 
the properties of the polarons and the 
experimental realization of these phases.
\end{abstract}

%\keywords{}


\maketitle
%%%%%%%%%%%%%%%%%%%%%%%%%%%%%%%%%%%%%%
\section{Introduction}
In recent years the experimental control of trapped ultracold atoms has evolved to a level that allows to probe sophisticated many-body effects, in particular the regime of strong correlations.
%In recent years the experimental control of dilute ultracold atoms 
%has evolved to a level that allows many sophisticated many-body effects 
%to be probed in the present laboratory. 
One of the most prominent achievements, the realization of the 
superfluid to Mott-insulator transition of bosonic atoms in an 
optical lattice \cite{Jaksch,MISF}, triggered both experimental and theoretical 
research into quantum phase transitions of ultracold atoms in optical lattices.
The advances in trapping and cooling techniques, as well as the manipulation 
of atom-atom interactions by Feshbach resonances \cite{feshbach}, created
the possibility to study many-body systems in the quantum degenerate 
regime in a widely tunable environment.
Of particular interest are those systems that resemble 
solid-state systems, for example,
mixtures of ultra-cold bosonic and fermionic atoms \cite{bfm},
which have recently become accessible through the development of 
sympathetic cooling \cite{sympathetic_cooling}.
On the theoretical side several phenomena have
been proposed in the literature like pairing of fermions \cite{pairing}, 
formation of composite particles \cite{composite}, spontaneous breaking 
of translational symmetry in optical lattices \cite{burnett,xCDW}. However,
the approach used in most of these studies, which relied on
integrating out the bosonic degrees of freedom and then using 
a mean-field approach to investigate many-body
states \cite{pairing}, becomes unreliable in the regime of 
strong interactions, in particular, in low-dimensional systems.
A non-perturbative and beyond-mean-field investigation is therefore 
of high interest.

In one-dimensional (1D) systems, on the other hand, 
many physical problems can be
solved exactly by various mathematical methods because of the 
restricted phase space and higher order geometric symmetry.
One of the most successful examples is the Luttinger liquid model
for 1D electron systems \cite{review}. The
pioneering work of Tomonaga \cite{tomonaga}, Luttinger \cite{luttinger} and
Haldane \cite{haldane,haldane_bf} along with many others has produced an
essentially complete understanding of the low energy physics
of these systems
(within the so called bosonization approach).
Recently these theoretical methods were also applied
to systems of ultra-cold atoms \cite{cazalilla,ho,mathey}, 
where the bosonic or fermionic atoms are confined in
a highly enlongated magnetic-optical trap
potential and approach an ideal 1D or quasi-1D system.
Such 1D or quasi-1D enlongated trap potentials can be easily prepared either 
in a traditional magnetic-optical trap \cite{1D_tradition}, 
magnetic waveguides on microchips \cite{microchip}, or in
an anisotropic optical lattice \cite{1Doptical_lattice}.
However, since there is no 
true long-range order in a homogeneous 1D system
 in the thermodynamic limit, the
one-dimensional ``quantum phases'' we will refer to in this paper are
actually understood in 
the sense of quasi-long-range order (QLRO), i.e. the correlation function of
a given order parameter ($O(x)$) has an algebraic decay at zero temperature,
$\langle O(x)O(0)\rangle\sim |x|^{-2+\alpha}$, where
$\lav\cdots\rav$ is the ground state average, $x$ is the distance, and $\alpha$
is the scaling exponent associated with that order parameter.
As a result, the dominant quantum phase is determined by the 
largest scaling exponent $\alpha >0$
(i.e. the slowest algebraic decay of the correlation functions of 
a given order parameter). 
%It is generally believed that these 
%QLRO phases can be stablized by the inter-chain coupling 
%and become a true long-range order via crossing to higher dimensional systems.
%
%
 At the phase boundaries that appear in the phase diagrams
 in this paper, the scaling exponent of the corresponding
 type of order becomes positive, i.e. the corresponding
 susceptibility switches from finite to divergent.

In this paper we extend and elaborate on our previous work \cite{mathey} on one-dimensional (1D) Bose-Fermi mixtures.
We give a detailed derivation of the effective low-energy Hamiltonian 
of a 1D BFM within a bosonization approach \cite{haldane_bf,cazalilla}. 
After diagonalizing this Hamiltonian we calculate the long-distance 
behavior of the single particle correlation functions.
We introduce a fermionic polaron ($f$-P) operator, 
constructed out of a fermionic operator 
with a screening cloud of bosonic atoms, and
 determine the single-particle correlation function, 
 as well as the 
correlation functions of various order parameters, which are
  used to construct the phase diagram.
%Motivated by the properties of the phase diagram,
%We show that 
%we introduce a fermionic polaron ($f$-P) operator, 
%constructed out of a fermionic operator 
%with a screening cloud of bosonic atoms in a BFM.
%We argue that this type of operator
%
% always
%has a slower decaying correlation function than the bare fermionic operator. 
%should be looked at as the underlying
% elementary operator, with which  
%Using the same polaronic operator we can also construct
%the pairing operator composed of two polarons (leading to fermionic polaron 
%pairing, $f$-PP) should be constructed.
%and show that it has a larger 
%(and positive)
%scaling exponent $\alpha$ (i.e. slower decaying correlation function) 
%then the pairing operator of bare fermions.
For a BFM of one species of fermions with one species of bosons,
we find a charge density wave (CDW) phase that competes with
the pairing phase of fermionic polarons 
 ($f$-PP),
%The CDW order parameter can also be considered to be constructed 
% of polarons (albeit that its properties are unaffected),
 giving a complete description of the system as a Luttinger liquid of polarons.
%
% as well as 
%a boson-fermion bound state
%(BFB) for commensurate filling. 
%can also compete with the polaron pairing phases in a different region
%of parameters. 
When the boson-fermion interaction strength is stronger than a critical value,
we observe the Wentzel-Bardeen instability (WB) region, where the BFM undergoes
phase separation (PS) for repulsive interaction or collapse (CL) for
 attractive interaction.

We also study several special cases when the density of the fermions
 or the bosons is commensurate with the
 lattice period.
 The low energy effective
 Hamiltonian of these cases
 cannot be diagonalized but instead has to be studied by using
  a renormalization group (RG) method.
 We obtain a charge gapped phase for half-filling of the fermions
 and for unit filling of the bosons, and study how these phases 
 are affected by the presence of the other species.

We also apply the bosonization method in an analogous approach to spinful fermions
mixed with a single species of bosonic atoms. 
%Experimentally 
%the two pseudo-spin states of fermions can be prepared either by using two
%different kinds of fermionic atoms or by the same type of atoms in different 
%hyperfine states. For simplicity, in this paper we consider only the
%latter case, where a global $SU(2)$ symmetry between the two spin states
%of fermions is assumed. Using the same bosonization approach,
We find a rich quantum phase diagram, 
%for spinful BFMs, 
including spin/charge density waves (SDW/CDW), triplet/singlet 
fermionic polaron pairing, and a regime
 showing the Wentzel-Bardeen instability.
%Due to 
The similarity with the known phase diagram of 
1D interacting electronic systems
\cite{review}, again suggests that a 1D BFM is best understood 
as a Luttinger liquid
of polarons.
% which has not been proposed before in the literature.
%
%We believe the concepts of polarons may be also important in higher 
%dimensional BFM, even though an exact solution is not available 
%in general.
%Therefore, for completeness and conceptual understanding, 
We
also discuss
the close relationship between the polarons constructed in the
bosonization approach in this paper and the canonical polaron transformation 
usually applied in solid state physics.

This paper is organized as follows: In Section \ref{Bos_Ham}, we present the 
microscopic description of a BFM and its bosonized representation.
In Section \ref{scaling}, we calculate the scaling exponents of single 
particle correlation functions as well as the correlation functions of 
various order parameters. Quantum phase diagrams obtained by comparing
these different scaling exponents are presented in the anti-adiabatic regime,
i.e. the phonon velocity is much larger than the Fermi velocity.
In Section \ref{other_parameter}, 
we present the results of 
%other parameter regimes,
%including half-filling in the fermionic sector, 
 two 
 commensurate filling
regimes:  
half-filling of fermions, and unit filling of bosons.
%(i.e. boson density is comparable to fermion density), 
%unit filling in the bosonic sector and
%the adiabatic regime.
In Section \ref{spinful}, we 
%show the bosonization approach to 
study a BFM of
spinful fermions with $SU(2)$-symmetry and determine the rich phase diagram 
of this system.
In Section \ref{discussion}, we discuss 
 questions regarding 
the experimental realization of the quantum 
phase diagrams presented in this paper, and 
 we summarize our results 
in
 Section \ref{conclusion}.
In Appendix \ref{A_scaling} we give a more technical
 derivation of the scaling exponents of various operators.
  In Appendix \ref{A_polaron} we discuss the notion of 
polarons used in this paper, and compare
 it to the those used in solid state
 systems. The effects of a local impurity 
in the system are discussed in Appendix \ref{impurity}.
%
%
%
%
%%
%%%%%%%%%%%%%%%%%%%%%
\section{Bosonized Hamiltonian and Exact Diagonalization}
\label{Bos_Ham}
%In this section, we first consider a mixture of spinless 
%(more precisely, a spin polarized) fermionic atoms with 
%a single component bosonic atoms in a 1D optical lattice (created by
%highly anisotropic 2D or 3D standing wave laser).
%In Subsection II.A we first derive and discuss the effective 
%low-energy Hamiltonian in bosonization representation. 
%In Subsection II.B we diagonalize the Hamiltonian.

%In this section and the next Section, we first derive the microscopic
%theory and the calculated quantum phase diagrams of a 1D BFM 
%in the anti-adiabatic regime (i.e. phonon velocity is much larger than 
%Fermi velocity) and high boson filling regime ($\nu_b\gg\nu_f$)
%without any commensurate filling.
%
%
In this section we derive and discuss
 the effective low energy Hamiltonian of a 1D BFM in the limit
 of large bosonic filling ($\nu_b \gg \nu_f$, where $\nu_{b/f}$ is the
 bosonic/fermionic filling fraction) and fast bosons 
 ($v_b\gg v_f$, where $v_b$ is the
 bosonic phonon velocity, and $v_f$ is the Fermi velocity).
We also mention how this 
 study can be extended to 
 the regime of comparable velocities ($v_f\sim v_b$).
%For this limit, 
%we derive  a description of the system which is diagonalizable, and, furthermore,
% the effective parameters appearing in the effective Hamiltonian
% can be related to the experimentally accessible parameters, such as the lattice
% strengths of the optical lattice, the scattering lengths and the densities
% of the bosonic and fermionic atoms.
%The main results of this section
% are the effective Hamiltonian, which 
%we
%present in momentum space representation, Eq.
% (\ref{Ham_modes}), and in real space (\ref{H_eff}), 
% the expressions for the parameters in this Hamiltonian, 
%(\ref{tbf}), (\ref{Ub}), and (\ref{Ubf}), and
% the Eqs. (\ref{vf}), (\ref{g}), (\ref{G}), (\ref{vb}) and (\ref{Kb}), and
% the Hamiltonian in diagonalized form (Section \ref{Ham_diag}).
%We note that the generic form of the effective
% Hamiltonian (\ref{H_eff}) extends beyond the
% parameter regime assumed in this section. 
%If we
% drop the assumption of fast bosons ($v_b\gg v_f$), and rather assume these
% velocities to be comparable,
% a renormalization argument of the type presented in \cite{voit_phonon}
% shows that the resulting Hamiltonian can also be written as (\ref{H_eff}). 
% Then, however, the expressions for the effective parameters shown in
% this section would not hold anymore, and would need to be replaced by parameters
% resulting from the RG flow.

%
%
%
%
%
We consider a BFM in an anisotropic optical lattice, where the lattice
strengths for bosons ($b$) and fermions ($f$) can be expressed as 
$V_{b (f)}({\bf r}) = \sum_{\alpha=x,y,z}  V_{b (f), \alpha} 
\sin^2(k_l r_\alpha)$ with
$k_l = 2\pi/\lambda$ being the laser wavevector and $\lambda$ being 
the wavevector of the laser fields. 
Throughout this paper we will use $\lambda/2=1$ (which corresponds to the
 lattice constant of the optical lattice) as 
the natural length scale. 
In order to create an effectively 1D lattice,
we consider $V_{b(f), ||} \ll V_{b(f), \perp}$, where
$V_{b(f), ||}=V_{b (f), x}$ is the lattice strength along the 
longitudinal direction and $V_{b(f), \perp}=V_{b (f), y}=V_{b (f), z}$
is along the transverse direction, so that the single particle
tunneling between each 1D tube is strongly surpressed.
We note that independent tuning of the optical lattice strength for these 
two species of atoms can be achieved even when only a single pair
of lasers provides the standing beam in each direction.
This 
can be done due to the fact that
the strength of an optical lattice for a given atom
is proportional to $\Omega_R^2/(\nu-\Delta E)$ \cite{bec_book}, 
where $\Omega_R$
is the Rabi frequency (which is proportional to the laser intensity),
$\nu$ is the laser frequency, and $\Delta E$ is the resonance energy
of that atom --- therefore for two different species of atoms (different
$\Delta E$), their effective lattice strengths can be tuned independently
over a wide range, by simultaneously 
 changing the laser intensities via $\Omega_R$ and 
 the laser frequency $\nu$.

%%%%%%%%
\subsection{Hamiltonian}
For sufficiently strong optical potentials the
microscopic Hamiltonian is given by a single band Hubbard model
\cite{Jaksch}:
\begin{eqnarray}
H &=& - \sum_{\langle ij\rangle} \left( t_b b_i^{\dagger} b_j 
+ t_f f_i^{\dagger} f_j \right)
-\sum_i\left(\mu_f n_{f,i}+\mu_b n_{b,i}\right)\nonumber\\
& &+ \frac{U_{b}}{2} \sum_i  n_{b, i} (n_{b,i} - 1)
+U_{bf} \sum_i n_{b, i} n_{f, i},
\label{H_tot}
\end{eqnarray}
where $n_{b,i}\equiv b^\dagger_i b^{}_i$ and $n_{f,i}
\equiv f^\dagger_i f^{}_i$ are the boson and fermion density 
operators, and $\mu_{b/f}$
are their chemical potentials.  The tunneling amplitudes $t_{f/b}$,
and the particle interactions $U_{b}$ and 
$U_{bf}$ can be calculated from the lattice strengths and 
the $s$-wave scattering lengths:
\bea
t_{b (f)} &  = &  (4/\sqrt{\pi}) \bar{V}_{b (f) , \parallel}^{3/4} E_{b (f)} 
\exp(- 2 \bar{V}_{b (f), \parallel}^{1/2})\label{tbf}\\
U_{b} &  =  & (8/\pi)^{1/2}  (2 \pi a_{bb}/\lambda) E_b (\bar{V}_{b, \parallel} 
\bar{V}_{b, \perp}^2)^{1/4}\label{Ub}\\
U_{bf} & = & (8/\pi)^{1/2} (2 \pi a_{bf}/\lambda) E_{bf} (\bar{V}_{bf, \parallel} 
\bar{V}_{bf, \perp}^2)^{1/4}\label{Ubf} 
\eea
We defined $\bar{V}_{b (f), \parallel/\perp}=V_{b (f), \parallel/\perp}/E_{b (f)}$ and $\bar{V}_{bf, \parallel (\perp)} =  
(4 \bar{V}_{b, \parallel (\perp)} \bar{V}_{f, \parallel (\perp)} )
/((\bar{V}_{b, \parallel (\perp)}^{1/2} + \bar{V}_{f, \parallel 
(\perp)}^{1/2})^2)$. $E_{b (f)} = (2\pi/\lambda)^2/2m_{b (f)}$ are the recoil energies of the two atomic species, 
 and $E_{bf}$ is given by $E_{bf}=(m_b+m_f)/4m_bm_f(2\pi/\lambda)^2$. 
%(For simplicity, we assume the mass of bosonic and fermionic atoms 
%are the same.)
%
%

For the convenience of the subsequent discussion we also give the Hamiltonian
in momentum space:
\bea\label{Ham_mom}
H & = & H_b^{(0)} + H_f^{(0)} + H_{bf}
\eea
Here, $H_b^{(0)}$ refers to the bosonic sector without the coupling to the 
fermions, and is given by:
\bea
H_b^{(0)} & = & \sum_k (\epsilon_{b,k}-\mu_b) b_k^\dagger b_k +
\frac{U_b}{2L} \sum \rho_{b,k}^\dagger\rho_{b,k}
\eea
$L$ is the 1D system length.
 $\epsilon_{b,k}$ is given by $\epsilon_{b,k} = -2 t_b \cos k$, and
 $\rho_{b,k}$ is defined as $\rho_{b,k}=\sum_p b_{p+k}^\dagger b_{p}$.
The free 
 fermionic sector - without the coupling to the bosons - is given by:
\bea
H_f^{(0)} & = & \sum_k (\epsilon_{f,k}-\mu_f) f_k^\dagger f_k
\eea
 where $\epsilon_{f,k}$ is given by $\epsilon_{f,k}=-2t_f\cos k$.
The interaction between these two sectors can be written as:
\bea
H_{bf} & = & \frac{U_{bf}}{L} \sum_{k} \rho_{f,k}^\dagger\rho_{b,k}
\eea
 with $\rho_{f,k}=\sum_p f_{p+k}^\dagger f_p$.
%
%
%%
%
%
%Before deriving the full effective Hamiltonian in bosonized form, 
%we consider it to be instructive to derive the equivalent
%low energy Hamiltonian 
%for each term of Eq. (\ref{Ham_mom}) individually.
%in terms of the elementary excitations of
%bosons and fermions. 
%
%
%
%

\subsection{Low energy effective Hamiltonian}
 To determine the effective low energy Hamiltonian, 
 we first  consider the
 effective Hamiltonian of 
 weakly interacting bosons, $H_b^{(0)}$, 
 without coupling to the fermions.
Following the standard Bogoliubov transformation 
\cite{bec_book,bogoliubove}, 
we obtain the following
effective low-energy Hamiltonian
for the bosonic sector: 
\begin{eqnarray}
H_b^{(0)} & \rightarrow & \sum_{k \neq 0} \omega_{b,k} \beta_{k}^{\dagger} \beta_{k},
\label{eq:H_b^0}
\end{eqnarray}
where the $\beta_k$ are the Bogoliubov phonon operators and
$\omega_{b,k} = \sqrt{(\epsilon_{b,k} - \epsilon_{b,0})
(\epsilon_{b,k} - \epsilon_{b,0} + 2 U_b \nu_b)}$ is the 
phonon dispersion.
% with the band energy $\epsilon_{b,k} = - 2 t_b \cos k$.
The phonon operators $\beta_k$ are related to the original boson
operators by $\beta_k=u_kb_k-v_k b_{-k}^\dagger$, where $u_k^2=1+v_k^2
=(\epsilon_{b,k}-\epsilon_{b,0}+U_b\nu_b+\omega_{b,k})/2\omega_{b,k}$
and $u_k v_k=-U_b\nu_b/2\omega_{b,k}$ are the coefficients of the Bogoliubov
transformation \cite{bec_book}.
We note that, although this tranformation is considered as 
a treatment of the quadratic fluctuations around a 
meanfield approximation, it is still
applicable to 1D systems where no true condensate 
exists even at zero temperature (see [\onlinecite{castin}]). 
%This is because,
In the long wavelength limit, we have
$\omega_{b,k}=v_b|k|$, where $v_b=\sqrt{2 t_b U_b \nu_b}$
is the phonon velocity. 
%One can easily obtain the same results by
%starting from a bosonization approach \cite{cazalilla}, in the
% weakly interacting limit.

For the fermionic sector we proceed along the lines of
the Luttinger liquid formalism, established in solid state physics 
 %in 
\cite{tomonaga,luttinger,haldane,review}. We linearize the 
noninteracting fermion band energy around the two Fermi points 
(at $\pm k_f$, with $k_f$ being the Fermi wavevector, $k_f=\pi\nu_{f}$), and
split the fermion operator into a right ($R$) and a left ($L$) moving channel.
 One can show that such a 
 linearized band Luttinger model \cite{luttinger}
is effectively equivalent to a bilinear bosonic 
Hamiltonian 
%representation 
in the 
low energy limit:
\begin{eqnarray}
H_f^{(0)} & \rightarrow & \sum_{k\neq 0} v_f |k| B^{\dagger}_k B_k,
\label{eq:H_f^0}
\end{eqnarray}
where $B_k$ is a bosonic (density) operator defined as 
$B_k\equiv i\sqrt{\frac{2\pi}{kL}}\sum_p f^\dagger_{R,k+p}f^{}_{R,p}$
for $k>0$ and $B_k\equiv -i\sqrt{\frac{2\pi}{|k|L}}\sum_p 
f^\dagger_{L,k+p}f^{}_{L,p}$ for $k<0$. $f_{R/L,k}$ are the right/left movers and 
\bea
v_f & = & 2t_f\sin(k_f)\label{vf}
\eea
 is the Fermi velocity.

Finally we will address the interaction between the bosonic
  and the fermionic atoms.
In the low-energy limit, there are two kinds of scattering to be
considered. One is the scattering by exchanging small
momentum in both the fermionic and the bosonic sector.
The other one is the scattering by exchanging
a momentum of $2 k_f$ to reverse the direction of a fermion moving in 1D.
For the first term,
only the long wavelength density fluctuations are important so
the corresponding interaction can be expressed  
by the same transformation used above:
\begin{eqnarray}
H_{bf}^{(1)} & \rightarrow & \sum_{k\neq 0} g |k| (\beta_k^{\dagger} 
+ \beta_{-k})(B_k + B_{-k}^{\dagger}).
\label{eq:H_bf^1}
\end{eqnarray}
where $g$ and $K_b$ are given by  
\bea
g & = & U_{bf} \sqrt{K_b}/2\pi\label{g}\\
K_b & = & \pi\sqrt{2t_b\nu_b/U_b} 
\eea
$K_b$ is the Luttinger parameter for bosons as 
will be discussed further in the next section.

%However, in one-dimensional fermionic systems, 
For the second case,
low energy excitations
 occur in the backward scattering channel, 
%where the momentum transfer
%is $2k_f$
 and therefore cannot be included in 
Eq. (\ref{eq:H_bf^1}), shown above.
Such a backward scattering term, in general, leads to a
non-linear and  
non-diagonalizable term in the Hamiltonian.
We therefore consider a certain parameter regime in which
 the backward scattering term becomes tractable.
%
%the effective Hamiltonian
%shown above. Such a backward scattering term, however, does not lead to an exactly solvable Hamiltonian
%in general and therefore we choose to consider a certain limit instead to 
%obtain a solvable 
%system.
In this and the next section (\ref{scaling}), we will consider 
a BFM with large bosonic filling (i.e. $\nu_b\gg \nu_f$), and 
 fast boson limit
 (i.e. $v_b\gg v_f$). 
 Experimentally, this limit can be achieved 
 by choosing $t_b\gg t_f$, and $\nu_b \approx 1 - 3$, for typical atomic interactions.  
For simplicity,
we will also assume that both the 
bosonic and the  
fermionic filling fraction are not commensurate to the lattice period 
or to each other.
% which would give rise 
% to additional non-linear terms.
%so that
%no Umklapp scattering is relevant here. 
We will show that the Hamiltonian of such a system can be solved exactly
even in the presence of backward scattering.
% and we will demonstrate 
%the physics of a BFM that can be tracted from this exact solution.

%In Section \ref{other_parameter} we will relax the above assumptions
% and use renormalization group analysis to investigate the
%possible new phases. 
%
%
%
%
%
In the limit of large bosonic filling and fast boson 
velocity, it is easy to see that 
the backward scattering of fermions is mainly provided by
the forward scattering of bosons (because $2k_f=2\pi\nu_f\ll 
2k_b=2\pi\nu_b$), so that the effective interaction 
Hamiltonian in this channel can be written as
\begin{eqnarray}
H_{bf}^{(2)} & \rightarrow & \frac{1}{\sqrt{L}}  
\sum_{|k|\sim 2 k_f} g_{2 k_f}  (\beta_k^{\dagger} + \beta_{-k})
\sum_p f^\dagger_{p+k} f^{}_p
\label{eq:interaction}
\end{eqnarray}
%
%
%
%
%
%where $n_k=\sum_p f^\dagger_{p+k} f^{}_p$ is the density operator and
where the coupling is given by 
$g_{k}=U_{bf}[\nu_b(\epsilon_{b,k}
-\epsilon_{b,0})/2\pi\omega_{b,k}]^{1/2}$. 
%$L$ is the 1D system length.
The next step is to 
separate the non-interacting phonon field, Eq. (\ref{eq:H_b^0}), into
 a low energy and high energy part, 
$H=\sum_{|k|<k_f}\omega_{k}\beta_k^\dagger\beta_k+\sum_{|k|>k_f}\omega_k
\beta_k^\dagger\beta_k$, and then integrate out the latter for $|k|\sim 2k_f$, 
together with the backward scattering term in Eq. (\ref{eq:interaction}).
Since we assume $v_b \gg v_f$, we can obtain an effective interaction
between fermions within an instaneous approximation. 
After some algebra, we can obtain such an effective fermion-fermion
 interaction to be:
%This gives a term 
%of the following form:
%
\begin{eqnarray}
H_f^{(1)}&=&-\frac{1}{L} \frac{g_{2 k_f}^2}{\omega_{2 k_f}} 
\sum_{|k| \sim 2 k_f} \sum_p f^\dagger_{p} f^{}_{p+k}
\sum_q f^\dagger_{q+k} f^{}_q
\nonumber\\
&=&\frac{1}{L} \frac{g_{2 k_f}^2}{\omega_{2 k_f}} \sum_{\alpha=\pm}
\sum_{k\sim 2\alpha  k_f} \sum_{p,q\sim -\alpha k_f} f^\dagger_{p} f^{}_q
f^\dagger_{q+k} f^{}_{p+k}
\nonumber\\
&=&\frac{1}{L} \frac{g_{2 k_f}^2}{\omega_{2 k_f}} 
\sum_{k'\sim 0} \left[n_{R,k'}n_{L,k'}+ n_{L,k'}n_{R,k'}\right]
\label{eq:H_f^20}
\end{eqnarray}
where $n_{R/L,k}\equiv \sum_p f^\dagger_{R/L,p+k} f^{}_{R/L,p}$
is the density operator for the right/left moving electrons.
We therefore find that the original backward scattering obtained by 
integrating out the boson field now becomes a forward scattering term 
with {\it repulsive} interaction between the left and right movers.
Therefore we can apply the bosonization approach for Luttinger
liquids used in Eq. (\ref{eq:H_f^0}) and obtain
\begin{eqnarray}
H_{f}^{(1)} & = & \sum_{k\sim 0} G |k| 
(B_{k} B_{-k} + B_{k}^{\dagger} B_{-k}^{\dagger} )
\label{eq:H_f^2}
\end{eqnarray}
where 
\bea
G & = &\frac{g_{2 k_f}^2}{\omega_{2 k_f}}.\label{G} 
\eea
Therefore the sum of the terms given in Eq. (\ref{eq:H_b^0}), (\ref{eq:H_f^0}),
(\ref{eq:H_bf^1}) and (\ref{eq:H_f^2}) is our final effective low energy 
Hamiltonian within the limit of large bosonic filling and fast boson velocity.
The total Hamiltonian then becomes:
\bea
H & = & 
 \sum_{k \neq 0} \omega_{b,k} \beta_{k}^{\dagger} \beta_{k}
 +\sum_{k\neq 0} v_f |k| B^{\dagger}_k B_k\nonumber\\
& &+\sum_{k\neq 0} g |k| (\beta_k^{\dagger} 
+ \beta_{-k})(B_k + B_{-k}^{\dagger})\nonumber\\
& & + \sum_{k\sim 0} G |k| 
(B_{k} B_{-k} + B_{k}^{\dagger} B_{-k}^{\dagger} ),
\label{Ham_modes}
\eea
%
%
%
%
%This Hamiltonian 
which is bilinear in two different
bosonic operators 
%(one is the Bogoliubov phonon operator $\beta_k$ from the 
%bosonic 
%excitations and the other
%one is the Luttinger boson $B_k$ from excitations of the fermions)
and therefore can be diagonalized exactly. 
For the convenience of later discussion and calculating the correlation
functions, in the next section we will use Haldane's 
 bosonization representation to express
the effective Hamiltonian shown above.

%However,
%for the sake of clarity and the convenience of calculating
%correlation functions, below we will also show the bosonization representation
%suggested by Haldane \cite{haldane} to treat bosons 
%and fermions on equal footing.

%%%%%%%%%%%%%%%%%%%%%
\subsection{Effective Hamiltonian in Haldane's bosonization representation}% and exact diagonalization}
We now bosonize both fermions and bosons on equal footing  
by introducing the phase and density fluctuation operators 
\cite{haldane_bf,cazalilla} through the definitions:
\bea
f(x)&=&\left[\nu_f+\Pi_f\right]^{1/2}\sum_{m=-\infty}^{\infty}
e^{(2m+1)i\Theta_f}e^{i\Phi_f}
\label{f_bosonization}
\\
b(x)& =&\left[\nu_b+\Pi_b\right]^{1/2}\sum_{m=-\infty}^{\infty}
e^{2mi\Theta_b}e^{i\Phi_b}
\label{b_bosonization}
\eea
where we use $x$ to denote  a continuous coordinate if there is 
no lattice potential
in the longitudinal direction or a discrete coordinate in the presence 
of a lattice. $\Pi_{b/f}(x)$ and $\Phi_{b/f}(x)$ are respectively
the density and phase fluctuations with $\Theta_{b/f}\equiv
\pi\nu_{b/f}x+\pi\int^x dy\Pi_{b/f}(y)$ accounting for the discreteness
of atoms along the 1D direction.
The density and phase fluctuations satisfy the commutation relations:
\begin{eqnarray}
\left[\Pi_{b/f}(x),\Phi_{b/f}(x')\right]=i\delta(x-x').
\label{commutator}
\end{eqnarray}
The advantage of using the density-phase fluctuation representation
for boson-fermion mixture is that it treats the fermions and bosons
in the same way. 
We do not have to use a Bogoliubov transformation
for the bosonic field and a Luttinger liquid formalism 
for fermions  separately as has been done
above.
% and usually in the literature. 
%It is therefore straightforward
%to show that 
As has been established in the literature (e.g. \cite{haldane_bf,cazalilla}),
the effective low energy Hamiltonian
of the bosonic and the fermionic sector can be written 
 in terms of the fields $\Pi_{b/f}$ and $\Phi_{b,f}$ as: 
\bea
H_b^{(0)} & = & \frac{v_b}{2}\int dx
\left[\frac{K_b}{\pi}
\left(\partial_x\Phi_b\right)^2+\frac{\pi}{K_b}\Pi_b^2\right]
\label{H_b^0}
\\
H_f^{(0)} & = & \frac{v_f}{2}\int dx
\left[\frac{K_f}{\pi}
\left(\partial_x\Phi_f\right)^2+\frac{\pi}{K_f}\Pi_f^2\right],
\label{H_f^0}
\eea
where we have $K_f=1$ due to the absence of $s$-wave short-range
interaction between fermionic atoms.
Note that 
%in the presence of a lattice potential in the 
%longitudinal direction, 
for intermediate values of
 the bosonic interaction $U_b$,
the phonon velocity $v_b$ and the 
Luttinger exponent $K_b$
are renormalized by higher order terms. They can be obtained
by solving the model exactly via Bethe-ansatz and
 are found to be very well approximated by
\bea
v_b& =& \frac{\nu_s}{m^\ast_b}\sqrt{\gamma}
\left(1-\frac{\sqrt{\gamma}}{2\pi}
\right)^{1/2}\label{vb}
\eea
 and 
\bea
K_b &=
&\frac{\pi}{\sqrt{\gamma}}\left(1-\frac{\sqrt{\gamma}}{2\pi}
\right)^{-1/2},\label{Kb}
\eea
%
%
%
%%
% in the weakly interacting limit, 
where $m^\ast_b$ is the effective boson mass in the
presence of the lattice potential \cite{stringari} and 
$\nu_s \sim \nu_b$ is the superfluid fraction.
$\gamma\equiv m^\ast_b U_b/\nu_s$ is a dimensionless parameter, characterizing
the interaction strength of the 1D boson system.
To leading order in the interaction (in the 
weakly interacting 
limit $\gamma\rightarrow 0$),
$v_b$ and $K_b$ are the same as before.
%
%

%Comparing Eqs. (\ref{H_b^0})-(\ref{H_f^0})
%with Eqs. (\ref{eq:H_b^0})-(\ref{eq:H_f^0}), we find that 

The 
density and phase fluctuations,
 defined in Eqs. (\ref{f_bosonization}) and 
 (\ref{b_bosonization}), can be related to the bosonic 
operators $\beta_k$ and $B_k$ by (see \cite{cazalilla})
\begin{eqnarray}
\Pi_b(x) & = & \frac{1}{2\pi}\sum_{k\neq 0}
\sqrt{\frac{2\pi K_b}{L|k|}}\,{\rm sgn}(k)e^{ikx}
\left(\beta_{k}+\beta_{-k}^\dagger\right)
\label{Theta}
\\
\Phi_b(x) & = & \frac{1}{2}\sum_{k\neq 0}
\sqrt{\frac{2\pi}{L|k| K_b}}\,{\rm sgn}(k)
e^{ikx}\left(\beta_{k}-\beta_{-k}^\dagger\right),
\label{Phi}
\\
\Pi_f(x) & = &\frac{1}{2\pi}\sum_{k\neq 0}
\sqrt{\frac{2\pi K_f}{L|k|}}\,{\rm sgn}(k)
e^{ikx}\left(B_{k}+B_{-k}^\dagger\right)
\label{Theta_F}
\\
\Phi_f(x) & = & \frac{1}{2\pi}\sum_{k\neq 0}
\sqrt{\frac{2\pi}{L|k| K_f}}\,{\rm sgn}(k)
e^{ikx}\left(B_{k}-B_{-k}^\dagger\right).
\label{Phi_F}
\end{eqnarray}
which can also be seen by
 comparing Eqs. (\ref{H_b^0})-(\ref{H_f^0})
 with Eqs. (\ref{eq:H_b^0})-(\ref{eq:H_f^0})

The long wavelength limit of the boson-fermion interaction and
the backward scattering Hamiltonian shown in Eqs. 
(\ref{eq:H_bf^1}) and (\ref{eq:H_f^2}) can also be 
expressed in terms of the density and phase fluctuations
 operators as follows:
\begin{eqnarray}
H_{bf}^{(1)} &=&  U_{bf}\int dx\, \Pi_b\,\Pi_f
\label{H_bf^1}
\\
H_{f}^{(1)} &=& \frac{2G}{2\pi}\int dx
\left[\pi^2\Pi_f^2-(\partial_x\Phi_f)^2\right].
\label{H_f^2}
\end{eqnarray}
Therefore the total low-energy effective Hamiltonian is given by
the sum of Eqs. (\ref{H_b^0}), (\ref{H_f^0}), (\ref{H_bf^1}),
and (\ref{H_f^2}):
\bea
H_{\rm eff}&=&H_f^{(0)}+H_b^{(0)}+H_{bf}^{(1)}+H_f^{(1)}
\label{H_eff}
\eea
which can be diagonalized easily as described 
 below.
%in Section \ref{Ham_diag}.
%
%
%
%
%

Before solving the effective Hamiltonian, we note that similar problems 
have
been investigated in the context of electron-phonon 
interactions in one-dimensional
solid state physics in the literature, both with \cite{voit_phonon,egger}
or without \cite{Engelsberg,meden} backward scattering of electrons.
%However, 
%
In 
%some of 
 the previous studies, 
the backward scattering of fermions
was treated in bosonized form before integrating out the phonon field.
The resulting effective interaction obtained after integrating out the high
energy phonon field is of the form \cite{voit_phonon}
\bea
H_f^{(1)}\propto \int\!\!\! \int d^2\bfr d^2\bfr' D_0(\bfr-\bfr')
\cos\left[2\Theta_f(\bfr)\right]\cos\left[2\Theta_f(\bfr')\right],
\nonumber\\
\label{H_f^2_boson}
\eea
where $\bfr=(x,v_f \tau)$ is a space vector in 1+1 dimension and 
$D_0(\bfr-\bfr')$ is the phonon propagator. (Note that the symbol of 
density and phase fluctuations used in this paper and in
 Eq. (\ref{H_f^2_boson}) is different from Ref.
\cite{voit_phonon}.) 
%Without further investigation, it is hard to see
%how such Hamiltonian can become bilinear form when taking the
%instaneous approximation, i.e. 
The instantaneous approximation of the 
phonon propagator is given by
$D_0(\bfr-\bfr')\propto\delta(\bfr-\bfr')$. 
 However, 
%and
% with that the expression 
 Eq. (\ref{H_f^2_boson}) does not seem to reduce to a 
bilinear term as shown Eq. (\ref{H_f^2}). 
%
%
%
%
%
%Therefore various perturbative treatments including 
%renormalization group were used to analyze the effects of this
%additional interaction in Refs. \cite{voit_phonon,egger}.
%This seems contraditory to our simple results 
%shown in Eq. (\ref{H_f^2}).
Such superficial disagreement can be resolved 
by taking the normal ordered form of $\cos(2\Theta)$ carefully, 
as treated by Sankar in Ref. \cite{sankar}.
%
%, and one finds the
%same term (\ref{H_f^2}) from Eq. (\ref{H_f^2_boson}).
In our treatment we keep 
%Therefore the phononic effects in 1D electronic system studied previous
%in Refs. \cite{voit_phonon,egger} may need examed more carefully.
 the fermionic form after integrating out the phonon field and
rearranging the order of fermions before using bosonization,
 which is a technically simpler and unambiguous procedure.
%representation as we have done above can easily avoid such 
%problem and 
It shows 
 that within the instantaneous limit 
 %how
 the 1D BFM system (and the related 
electron-phonon system) can be described by a bilinear 
  bosonized Hamiltonian, 
 %in bosonized form, 
 which then can be diagonalized 
exactly, even in the presence of backward scattering.
%, if the phonon velocity
 %is much larger than the Fermi velocity.
%
%
%
%

We note that the generic form of the effective
 Hamiltonian (\ref{H_eff}) extends beyond the
 parameter regime assumed in this section. 
 If we
 drop the assumption of fast bosons ($v_b\gg v_f$), and rather assume these
 velocities to be comparable,
 a renormalization argument of the type presented in \cite{voit_phonon}
 shows that the resulting Hamiltonian can also be written as (\ref{H_eff}). 
 Then, however, the expressions for the effective parameters shown in
 this section would not hold anymore, and would 
need to be replaced by parameters
 resulting from the RG flow. 
%
%

%%%%%%%%%%%%%%%%%%%%%%%%%%%%%%%%%%%%%%%%%%%%%%
\subsection{Diagonalized Hamiltonian}\label{Ham_diag}
We now go on to diagonalize the full effective low energy 
Hamiltonian (Eq. (\ref{H_eff})) by using the following linear 
transformation \cite{Engelsberg}:
\begin{eqnarray}
\Pi_b=\delta_1\Pi_A+\delta_2\Pi_a, &&
\Phi_b=\epsilon_1\Phi_A+\epsilon_2\Phi_a,
\nonumber\\
\Pi_f= \beta_1\Pi_A+\beta_2\Pi_a, &&
\Phi_f= \gamma_1\Phi_A+\gamma_2\Phi_a,
\label{Transformation_Engelsberg}
\end{eqnarray}
where $\Pi_{A/a}$ and $\Phi_{A/a}$ are the density and phase operators
of the two eigenmodes, and the coefficients are given by:
\begin{eqnarray}
\begin{array}{ll}
\beta_1 =  e^\theta \sqrt{\frac{\tilde{v}_f}{v_A}} \cos \psi 
& \beta_2  =  e^\theta \sqrt{\frac{\tilde{v}_f}{v_a}} \sin \psi\\
\gamma_1  =  e^{-\theta} \sqrt{\frac{v_A}{\tilde{v}_f}} \cos \psi 
& \gamma_2   = e^{-\theta} \sqrt{\frac{v_a}{\tilde{v}_f}} \sin \psi\\
\delta_1 =  - e^\phi\sqrt{\frac{v_b}{v_A}} \sin \psi  
& \delta_2  =  e^\phi\sqrt{\frac{v_b}{v_a}} \cos \psi\\
\epsilon_1  =  - e^{-\phi}\sqrt{\frac{v_A}{v_b}} \sin \psi 
& \epsilon_2   =  e^{-\phi}\sqrt{\frac{v_a}{v_b}} \cos \psi.
\end{array}
\label{coefficients}
\end{eqnarray}
Here we have $e^\theta = 
((v_f - 2 G)/(v_f + 2 G))^{1/4}$,
$e^{\phi}=\sqrt{K_b}$, and
$\tan 2\psi = 4 \tilde{g} (v_b \tilde{v}_f)^{1/2}/(v_b^2 - \tilde{v}_f^2)$.
The diagonalized Hamiltonian 
\begin{eqnarray}
H&=&\frac{1}{2}\sum_{j=a,A}v_j\int dx\left[\pi\Pi_j(x)^2
+\frac{1}{\pi}\left(\partial_x\Phi_j(x)\right)^2\right]
\label{new_hamiltonian}
\end{eqnarray}
has the eigenmode velocities, $v_A$ and $v_a$, given by
\begin{eqnarray}
v_{a/A}^2 & = & \frac{1}{2} (v_b^2 + \tilde{v}_f^2) \pm \frac{1}{2} 
\sqrt{(v_b^2 - \tilde{v}_f^2)^2 +  16 \tilde{g}^2  v_b \tilde{v}_f},
\label{new_velocity}
\end{eqnarray}
where $\tilde{v}_f\equiv(v_f^2-4G^2)^{1/2}$ and $\tilde{g}\equiv
g\,e^\theta$, $e^{\theta}$ is given by $e^\theta = 
((v_f - 2 G)/(v_f + 2 G))^{1/4}$. 

From Eq. (\ref{new_velocity}) we note that when the fermion-phonon coupling $g$
(proportional to $U_{bf}$) becomes sufficiently strong, 
 %$4 \tilde{g}^2>\tilde{v}_f v_b$, 
 the eigenmode
 velocity $v_A$ becomes imaginary, indicating an 
instability of the system to phase separation (PS) or
collapse (CL), depending on the sign of $U_{bf}$ \cite{ho,instability}.
This is the so-called Wentzel-Bardeen instability, and has
 been studied before in 1D electron-phonon systems \cite{ho}
%
%, without
% the modification of $\tilde{g}$ and $\tilde{v}_f$
% due to the inclusion of the backscattering parameter $G$.

%However, to understand the nature of the many-body state 
%of BFM in the stability region
%we have to analyze the long distance behavior of the correlation functions
%as shown below.

%In the diagonalized Hamiltonian (\ref{new_hamiltonian}), 
%the effective parameters $K_A$ and $K_a$ are given by 
%\bea
%K_A & \equiv & K_\beta = \Big(K_{\gamma}^{-1} - 
%\frac{K_{\epsilon}}{K_{\gamma \epsilon}^2} \Big)^{-1}\\
%K_a & \equiv & K_\delta = \Big(K_{\epsilon}^{-1} - \frac{K_{\gamma}}
%{K_{\gamma \epsilon}^2} \Big)^{-1}
%\eea
%The rhs of these equations can be checked directly from the 
%definitions of the K-parameters. If we use the fields 
%$\Pi_A$/$\Theta_A$ and $\Pi_a$/$\Theta_a$ to construct fermion 
%and boson operators, respectively, we have obtained an effective 
%Luttinger liquid of polarons as we discuss in more detail below.

%
%
%
%%%%%%%%%%%%%%%%%%%%%%%%%%%%%%%%%%%%%%%
\section{Scaling exponents and quantum phase diagrams}
\label{scaling}
In this section, we will calculate the scaling exponents of various
quasi-long-range order parameters 
%and  their values 
to construct  
the quantum phase diagram.
The diagonalized form of the Hamiltonian, which we
 presented in Section \ref{Ham_diag}, allows an exact 
calculation of all correlation functions (see Appendix \ref{A_scaling}).
For the later discussion,
we define the following ``scaling exponents'' 
to mimic the ones in standard
single component bosonic \cite{cazalilla} or fermionic systems \cite{review}:
\bea
K_{\beta} &\equiv& \beta_1^2+\beta_2^2, \ \ \ 
K_{\gamma}^{-1} \equiv \gamma_1^2+\gamma_2^2
\\
K_{\delta} &\equiv& \delta_1^2+\delta_2^2, \ \ \ 
K_{\epsilon}^{-1} \equiv \epsilon_1^2+\epsilon_2^2
\\
K_{\beta\delta} &\equiv& \beta_1\delta_1+\beta_2\delta_2 \ \ \
K_{\gamma\epsilon}^{-1} \equiv  \gamma_1\epsilon_1+\gamma_2\epsilon_2.
\label{K}
\eea
Note that since $v_a>v_A$ according to Eq. (\ref{new_velocity}),
we have $K_{\beta\delta}<0$ and $K_{\gamma\epsilon}^{-1}>0$.
%Their physical meaning will be clarified in the later discussion.
%It is easy to see 
From Eq. (\ref{Transformation_Engelsberg}) it is apparent 
that $K_\beta$ and $K_\gamma^{-1}$
are the scaling exponent associated with the fermion density and phase 
fluctuations, respectively, and $K_\delta$ and $K_\epsilon^{-1}$
are associated with the boson density and phase fluctuations.
% respectively.
 $K_{\beta\delta}$ and $K^{-1}_{\gamma\epsilon}$ are related to  
fluctuations of mixed components,
 and appear in the scaling exponents of products of fermionic and
 bosonic operators. 
%
%
%
%%%%%%%%%%%
\subsection{Single particle correlation functions and polaron operators}
\label{single_p}
In this section
we consider the single-particle correlation functions of the system. 
For the bare bosonic and fermionic particles we find $\langle
b(x)b^\dagger(0)\rangle \sim |x|^{-\frac{1}{2}K_\epsilon^{-1}}$ and
$\langle f(x)f^\dagger(0)\rangle \sim \cos(k_f x)
|x|^{-\frac{1}{2}(K_\beta+K_\gamma^{-1})}$,
 as discussed in Appendix \ref{A_scaling}.
The scaling exponents that appear in  
 these correlation functions have been renormalized by the
 coupling between the bosons and the fermions.
 % Without that coupling, the parameter $K_\epsilon$ would reduce 
%  to $K_b$, the Luttinger parameter of the uncoupled
%  bosonic LL, and $K_\beta$ and $K_\gamma$ would both reduce to $1$, the
% value for a non-interacting Fermi gas.
% %
%  A key observation about the
% correlation functions of the BFM, 
% which sets these correlation functions
% apart from the corresponding correlation functions
%  of standard LL systems, is that the density fluctuations
%  and the phase fluctuations are now governed by different scaling exponents.
% For the fermions these are $K_\beta$ and $K_\gamma$, for the
% bosons  $K_\delta$ and $K_\epsilon$, and, in general, we have $K_\beta\neq K_\gamma$
%  and $K_\delta\neq K_\epsilon$.
%
 % 
%%The latter is similar but somewhat
%different from the standard Luttinger liquid form \cite{cazalilla,review},
%because $K_\beta\neq K_\gamma$ due to the appearance of
%bosonic degree of freedom.
%However, as we will show below these bare fermionic and bosonic operators 
%are not the most appropiate description of quasi-particles in 1D BFM. 

Motivated by 
 the polaronic effects in electron-phonon systems, 
%the study of the phase diagram (see Section \ref{PD} and 
%Appendix \ref{A_scaling}),
 we next consider a class of operators
 that seems to be most appropiate as the elementary operators of the system.
%, the
% polaron operators.
%, as motivated by
% the study of the phase diagram.
%
 In BFMs each atom will repel (attract) the atoms of the other species in 
its vicinity, due to their mutual interaction,
 resulting in a reduced (enhanced) density around that atom. 
%i.e. polaronic effects.
 To describe particles dressed
with  screening clouds of the other species
we introduce the following composite operators (polarons):
\begin{eqnarray}
\tilde{f}_\lambda(x) \equiv e^{-i\lambda\Phi_b(x)} f(x),
\label{f_polaron}
\\
\tilde{b}_\eta(x) \equiv e^{-i\eta\Phi_f(x)} b(x),
\label{b_polaron}
\end{eqnarray}
with $\lambda$ and $\eta$ being real numbers, describing the 
size of the screening cloud. 
%The physical reason of constructing these composed
%bosonic and fermionic operators in this certain  will be 
%critically discussed in Section \ref{discussion} in the
%context of standard polaron theory. 
%Basically, t
 The
correlation functions of these new fermionic and bosonic operators
can be calculated to be (cp. Appendix \ref{A_scaling})
\bea
\langle\tilde{f}_\lambda^{} (x)\tilde{f}^\dagger_\lambda(0)\rangle 
&\sim& \cos(k_f x)
|x|^{-\frac{1}{2}(K_\beta +K_\gamma^{-1}
-2\lambda K_{\gamma\epsilon}^{-1}+\lambda^2 K_\epsilon^{-1})}
\nonumber\\
\\
\langle\tilde{b}_\eta^{}(x)\tilde{b}^\dagger_\eta(0)\rangle &\sim&
|x|^{-\frac{1}{2}(K_\epsilon^{-1}+\eta^2 K_\gamma^{-1} -2\eta
K_{\gamma\epsilon}^{-1})}.
\eea
Treating $\lambda$ and $\eta$ as variational parameters,
we observe that the exponents of the
correlation functions are 
minimized (i.e. the correlation
 functions have the slowest decay at long distances)
 by taking
$\lambda_c=K_\epsilon/K_{\gamma\epsilon}$ and
$\eta_c=K_\gamma/K_{\gamma\epsilon}$.
To gain some insight into the quantity
 $\lambda_c$
%, which is the more relevant of the two parameters
 %for the parameter regime that we consider,
 we consider
%
%To understand the physical meaning of the values of $\lambda_c$ 
%and $\eta_c$
%, we can, for example, take 
%
the limit of weak
boson-fermion interaction
($U_{bf}\rightarrow 0$) and obtain approximately 
%$$
%
%The physical meaning of these parameters, $\lambda_c$ and
%$\eta_c$, can be understood in a simple way:
%In the limit of weak
%interactions we have 
%
$\lambda_c\sim U_{bf}/U_b$ (for further discussion, see Appendix \ref{A_polaron}). 
%and
%$\eta_c\to 2 U_{bf}/\pi v_b$. 
This result can be interpreted
by a simple density counting argument
 as follows:
we imagine a single fermionic atom interacting with $-\lambda$ bosons
 in its vicinity. The potential
 energy of such a configuration can be estimated
 as  $-U_{bf}\lambda + U_b \lambda^2/2$, 
 which is minimized by taking $\lambda=U_{bf}/U_b$.
%
%that a fermionic polaron locally suppresses a bosonic cloud of
%$\lambda_c$ particles, whereas a bosonic polaron 
%depletes the fermionic system by $\eta_c$ atoms. 
Since the polaronic
cloud size is not a well-defined quantum number, $\lambda_c$ and
$\eta_c$ need not be integers and depend on the boson-fermion 
interaction strength continuously.

Taking the optimal values of $\lambda_c$ and $\eta_c$
in Eqs. (\ref{f_polaron})-(\ref{b_polaron}) and using 
the following two identities:
$K_\beta^{-1}=K_{\gamma}^{-1} - K_{\epsilon}K_{\gamma \epsilon}^{-2}$ 
and $K_\delta^{-1} = K_{\epsilon}^{-1} -K_{\gamma}K_{\gamma \epsilon}^{-2}$, 
we can show that
\bea
\langle\tilde{f}_{\lambda_c}^{} (x)\tilde{f}^\dagger_{\lambda_c}(0)\rangle 
&\sim& \cos(k_f x)|x|^{-\frac{1}{2}(K_\beta +K_\beta^{-1})}
\label{f_scaling}
\\
\langle\tilde{b}_{\eta_c}^{}(x)\tilde{b}^\dagger_{\eta_c}(0)\rangle &\sim&
|x|^{-\frac{1}{2}K_\delta^{-1}}.
\label{b_scaling}
\eea
In other words, the long-distance (low energy) behavior of 
both the fermionic-polaron ($f$-P, defined 
in Eq. (\ref{f_polaron}) with $\lambda=\lambda_c$) and the 
bosonic-polaron ($b$-P, defined in  Eq. (\ref{b_polaron})
with $\eta=\eta_c$) are 
characterized by a single scaling parameter, $K_\beta$ and $K_\delta$, 
respectively, in contrast 
 to the scaling behavior of the bare atoms.
%
%Indeed, as we will show later that 
As we will see in the next section, 
the many body correlation functions of operators 
composed of such fermionic polarons also scale with the {\it same}
 single parameter,
$K_\beta$, leading to the conclusion that the 1D BFM system is best
understood as a Luttinger liquid of polarons,
instead of bare fermions.
%We point out that this is the first time that polaronic physics is
%explicitly developed in the context 
%of 1D BFM. Similar studies might also be
%applied to other systems of higher dimension or systems of interacting
% electrons and phonons.
%analogous to a regular Luttinger liquid description for
%1D electronic systems in solid state physics \cite{review}.

%--------------
%
%
%\begin{figure}
%\includegraphics[width=8cm]{PD_bare.eps}
%\caption{
%
%Ground state of a BFM with spinless fermions.
%$g$ is the longitudinal FP coupling and $G$ is the effective
%fermion-fermion interaction after integrating out $2k_f$ phonons.
%
%(a) 
%Regions of QLRO of $O_{CDW}$ and $O_{BFP}$.
%For larger values of $G$ we find that CDW ordering dominates, for larger 
%values of $g$ we find pairing of the bare fermions.
%However, in between these two regimes, we find a region in which neither of these
%two operators shows QLRO. 
%(b) In this diagram we additionally take operators of the type $O_{CP,n}$ into
% account, and plot the dominant order.
%These phases of composite pairing make up almost the entire 
%regime between CDW ordering and phase separation (PS).
% For both of these diagrams, we chose $v_b/v_f=5$ and $K_b=10$.
%}
%\label{phase_diag_BFP}
%\end{figure}
%
%
%
\begin{figure}
\includegraphics[width=8cm]{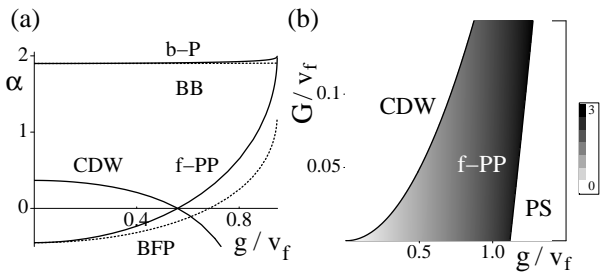}
\caption{
%
%Ground state of a BFM with spinless fermions.
%$g$ is the longitudinal FP coupling and $G$ is the effective
%fermion-fermion interaction after integrating out $2k_f$ phonons.
%
(a) Scaling exponents of various order parameters as a function
of longitudinal FP coupling $g$ for a BFM with spinless fermions.
Parameters are chosen to be: $v_b/v_f = 3$, $K_b = 5$ and $G/v_f = 0.1$.
Different curves correspond to the $2k_f$--CDW, $f$-polaron pairing 
field ($f$-PP), bare fermion pairing
field (BFP), $b$-polaron operator ($b$-P), and bare boson operator
(BB). Note that operators
constructed with polarons (i.e. $b$-P and $f$-PP) 
always have larger exponents than their counterparts
constructed with bare atoms. (b) Global phase diagram 
of the same system in terms of effective forward scattering between fermions,
$G$ and the fermion-phonon coupling $g$ for $v_b/v_f=5$
and $K_b=10$. Three different phases, CDW, $f$-PP, and phase separation
(PS) are dominant successively from weak to strong fermion-phonon 
coupling strength.}
\label{phase_diag_alpha}
\end{figure}
%-----------------
%
%
%
%
%
%%%%%%%%%%%%%%%%%%%%%%%%%%%%%
\subsection{Correlation functions of order parameters}% and Phase diagrams}
\label{PD}
%
%
%
%
%We will now determine the phase diagram
%of an incommensurate BFM in 1D.
%
%
%This will be done by
%comparing the scaling exponents of various order parameters by
%calculating the long-distance behavior of their correlation functions.
%
%%
%Rather than simply presenting the diagram in its 
%final form, we will first consider the order parameters that are important
% for a simple LL of spinless fermions. Next we extend that analysis by considering
% pairing of composite pairs, in which Cooper pairs of fermions bind an integer number
% of bosons, and finally we consider polaron pairing, which envelopes all of these
% pairing phases.

%
%Luttinger liquids are typically unstable to some kind of ordering at $T=0$. 
%What ordering will occur can be determined by finding the slowest 
%decaying order parameter at long distances
%
%As mentioned before, 
%In this paper, 
 It is well established that there is
 no true off-diagonal long range order in 1D systems, due
 to strong 
%the importance of 
 fluctuations.
Therefore, in this paper, 
%We are interested in the zero temperature quantum
%phases, which in 1D systems are defined in the sense of quasi-long range order,
%i.e. 
 the ground state is characterized by the
order parameter that has the slowest long distance decay
of the correlation function \cite{review}, i.e.
 in the sense of QLRO. Transforming to momentum-energy
space, this is equivalent to
finding the most divergent susceptibility in the low temperature limit:
if, at $T=0$, the  correlation function 
of a certain order parameter $O(x)$
decays as $\langle O(x) O^\dagger(0)
\rangle \sim 1/|x|^{2-\alpha}$ 
 for large $x$,
the finite $T$ susceptibility will diverge
as $\chi(T) \sim 1/T^{\alpha}$.
Here we defined the scaling exponent $\alpha$, which is
 positive for a quasi long range ordered state (i.e. $\chi(T)$ diverges for 
$T\rightarrow 0$), and negative if no such order is present.
% if the scaling exponent $\alpha$ is positive. 
%
%
%
 Across a phase boundary the susceptibility will switch from finite to divergent. 
Since we are interested here 
  in the limit of high phonon velocity ($v_b\gg v_f$),
only the fermionic degrees of freedom need to be considered for
the low-energy (long-distance) behavior of the correlation functions. 
 The bosonic sector is always assumed to be in a quasi-condensate state.
(We will 
 relax these assumptions below.)

%
%
%
%
%
%
%Inspired by the phase diagram of 
 Similar to 
a LL of spinless fermions, we
first consider the order parameters of the CDW phase, defined as
 $O_{CDW}=f^\dagger_{L}(x) f_{R}(x)$, and of the $p$-wave pairing
 phase, given by 
$O_{BFP}=f_{L}(x)f_{R}(x)$, where  the abbreviation BFP 
 stands for bare fermion pairing.
%, to contrast this type of pairing with the
% pairing of fermionic polarons that we discuss below.
The scaling exponents for the CDW and the BFP phase can be calculated 
(see Appendix \ref{A_scaling})
 to be:
\bea
\alpha_{CDW} & = & 2 -2 K_\beta\label{CDW_scaling}\\
\alpha_{BFP} & = & 2 -2 K^{-1}_\gamma
\eea
In Fig. \ref{phase_diag_alpha} (a), we show the calculated scaling 
exponents for CDW and BFP phases as a function of the long-wavelength
 fermion-phonon coupling strength $g$, which is proportional to $U_{bf}$.
The other parameters are given by 
 $K_b=5$, $v_b/v_f=3$, and $G/v_f=0.1$.
One can see that CDW ordering becomes dominant ($\alpha_{CDW}>0$) for
$g/v_f<0.55$, while bare fermion pairing is dominant ($\alpha_{BFP}>0$) when
 $g/v_f>0.65$. In the intermediate regime
($0.55<g/v_f<0.65$) none of these two phases have quasi long range order, and 
therefore one might conclude that in this regime  
%there is no  
%QLRO, so that
 the system displays a metallic phase. 
%like a non-interacting 1D fermionic system.
A similar result has been discussed in the context of 1D 
electron-phonon systems, 
\cite{voit_phonon, egger, martin}.

However, as mentioned in our earlier publication \cite{mathey}, 
 the above analysis
 is incomplete.  
An indication for this is given by 
 how a single impurity potential 
 affects the transport of the 1D BFM (see Appendix \ref{impurity}). 
Following the renormalization group analysis by 
Kane and Fisher \cite{kane}, we 
 show that a single weak impurity potential
 is relevant when $\alpha_{CDW}=2-2K_\beta>0$ and becomes irrelevant
 whenever the system is outside the CDW regime, 
i.e.  when $\alpha_{CDW}=2-2K_\beta<0$. 
%(Details of the calculation are
% given in Appendix \ref{impurity}).
This result is inconsistent with the previous 
CDW-metal-BFP scenario, 
%because a metallic state, 
because
 a metallic state would not be insensitive to an impurity potential in 1D.
 More precisely, the fact that the impurity potential becomes irrelevant
  even outside the BFP phase indicates that there is another superfluid-like
 ordering, which cannot be described by the pairing of bare fermions only.

\begin{figure}
\includegraphics[width=6cm]{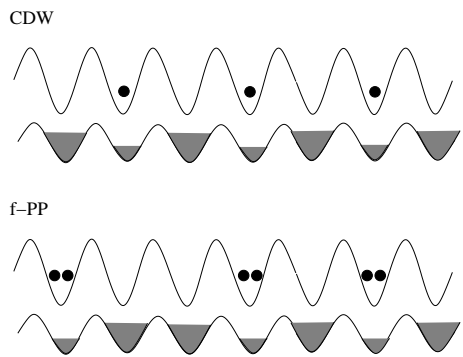}
\caption{
%
%Ground state of a BFM with spinless fermions.
%$g$ is the longitudinal FP coupling and $G$ is the effective
%fermion-fermion interaction after integrating out $2k_f$ phonons.
%(a) Scaling exponents of various order parameters as a function
Illustration of the two phases that occur 
in a BFM with spinless fermions, CDW and $f$-PP.
In the CDW phase the system develops a $2 k_f$
 density modulation in both the fermionic and the bosonic liquid.
 In the $f$-PP phase, fermionic polarons pair up and form a 
 superfluid state. 
}
\label{CDW_fPP}
\end{figure}

Above observation motivates a further search for the
appropiate order parameters. 
%In Appendix \ref{A_scaling} 
 %We consider
Among the many types of operators that can be considered, we find that
 the operator 
$O_{f-PP}=\tilde{f}_{L,\lambda_c}(x)\tilde{f}_{R,\lambda_c}(x)$, which describes
%
%
%Inspired by the above impurity study and the fermionic polarons  
%in the single particle correlation function, we define 
%a new order parameter, 
 a $p$-wave fermionic polaron pairing 
phase ($f$-PP), 
%defined as
%$O_{f-PP}=\tilde{f}_{L,\lambda_c}(x)\tilde{f}_{R,\lambda_c}(x)$.
%
shows dominant scaling outside of the CDW regime.
In Fig. \ref{CDW_fPP} we give an illustration of the two
 phases that occur in BFMs with spinless fermions.

%
%With these order parameters we find a phase diagram given in 
% Fig. \ref{phase_diag_BFP} (a). 
%The backscattering term (corresponding to $G$) tends to enhance CDW fluctuations 
% whereas the acoustic coupling ($g$) favors pairing.
%For a standard LL of spinless fermions, these two types of ordering would cover the 
%entire phase diagram. However, here we find that there is a regime
% where neither of these two operators develops QLRO.
%
%Because we have an additional field present - the bosons - there is a much larger
% number of types of order parameters that can be considered.
%Among these, the following class can develop dominant QLRO in our system: 
%$O_{cp,n}=f_{L}(x)f_{R}(x)(b^\dagger(x))^n$. These operators describe Cooper pairs
% to which $n$ bosons are bound, to create a composite pair (CP).
%In Fig. \ref{phase_diag_BFP} (b), we show the regions in which these operators
% develop dominant QLRO. 
%We note that almost the entire regime outside of CDW ordering
% shows pairing of some type of these composite pairs.
%
%These phases can be naturally interpolated by considering the pairing of polarons.
%
%{\it ....... more }

%
%
%
%
%For a BFM with spinless fermions 
%So we consider the following two 
%ordering instabilities: a Peierls-type instability towards CDW,
%whose order parameter is $O_{CDW}=f^\dagger_{L}(x) f_{R}(x)$, and 
%a p-wave $f$-polaron pairing with
%$O_{f-PP}=\tilde{f}_{L,\lambda_c}(x)\tilde{f}_{R,\lambda_c}(x)$.
%
%
%
%
%
%
%
%
%Using the results of bosonization, we can calculate the associated
%
As shown in Appendix \ref{A_scaling}, 
the scaling exponent of the fermionic polaron
  pairing phase can be calculated to be:
\bea
%\alpha_{CDW} &=& 2-2K_\beta
%\label{CDW_scaling}
%\\
\alpha_{f-PP} &=& 2-2\left[\lambda^2_cK_\epsilon^{-1}
+K_\gamma^{-1}-2\lambda_c K_{\gamma\epsilon}^{-1}\right]
\nonumber\\
&=&2-2K_\beta^{-1},
\label{f-PP_scaling}
\eea
where we used the same polarization parameter $\lambda_c$ as in the previous
 section, and the same algebraic relation between the scaling quantities as
 for the derivation of (\ref{f_scaling}). Note that the expression (\ref{f-PP_scaling})
is dual to the scaling of the CDW operator, Eq. (\ref{CDW_scaling}),
%
%
%Using the results of bosonization, we can calculate the associated
%
%The scaling exponent of this type of pairing phase can be calculated to be:
%
%
%
%
%
%
%\bea
%\alpha_{CDW} &=& 2-2K_\beta
%\label{CDW_scaling}
%\\
%\alpha_{f-PP} &=& 2-2\left[\lambda^2_cK_\epsilon^{-1}
%+K_\gamma^{-1}-2\lambda_c K_{\gamma\epsilon}^{-1}\right]
%\nonumber\\
%&=&2-2K_\beta^{-1},
%\label{f-PP_scaling}
%\eea
%
%
%
%
%
%
%where we used the same polarization parameter $\lambda_c$ as in the previous
% section, and the same algebraic relation between the scaling quantities as
% for the derivation of (\ref{f_scaling}). Note that the expression (\ref{f-PP_scaling})
%is dual to the scaling of the CDW operator, Eq. (\ref{CDW_scaling}).
%
%
%
%If we would cont
%
%Surprisingly, as mentioned above, 
 and, hence, both the CDW phase and the $f$-polaron
phase are governed by a {\it single} Luttinger parameter, $K_\beta$, which
also appears in the single particle correlation function of $f$-polarons
as shown in Eq. (\ref{f_scaling}). The fact that 
the scaling exponents (\ref{f_scaling}), (\ref{CDW_scaling})
and (\ref{f-PP_scaling}) 
%are the same as for 
are the ones of 
a Luttinger liquid of
1D fermion with parameter $K_\beta$ 
indicates that the 1D BFM system should be understood
as a Luttinger liquid of polarons.
%This interpretation is also reflected 
% in the corresponding operators, $\tilde{f}_{\lambda_c}$, 
% $O_{f-PP}=\tilde{f}_{L,\lambda_c}\tilde{f}_{R,\lambda_c}$,
% and $O_{CDW}=\tilde{f}^\dagger_{L,\lambda_c} \tilde{f}_{R,\lambda_c}$.
%For the latter 
We note that 
$\tilde{f}^\dagger_{L,\lambda_c} \tilde{f}_{R,\lambda_c}=f^\dagger_{L,\lambda_c} f_{R,\lambda_c}$,
 as can be seen from the definition of $\tilde{f}$, Eq. (\ref{f_polaron}).
% The operator $O_{CDW}$ has no net fermionic charge, and,
% 
Therefore, the scaling exponent of $O_{CDW}$
 is not affected by the screening
 cloud. 
In Fig. \ref{phase_diag_alpha}(a) we also  show the 
scaling exponent of 
the new type of  
 order parameter, $O_{f-PP}$.
We note that $\alpha_{f-PP}$  
is always
larger 
than $\alpha_{BFP}$, showing that BFP cannot be
the most dominant quantum phase in the whole parameter regime of 
a BFM. Furthermore, the scaling
exponents shown in Fig. \ref{phase_diag_alpha}(a) demonstrate that
divergencies of the CDW and $f$-PP susceptibilities are mutually exclusive
and cover the entire phase diagram 
up to the
%outside of
  phase separation (PS) regime, 
 which  can also been seen directly from 
 the scaling exponents in Eqs. 
(\ref{CDW_scaling}) and (\ref{f-PP_scaling}).
 This 
 explains why a single
 impurity can be
 irrelavant 
 when the system is in the regime between the CDW phase and the BFP phase,
 as discussed above.
%
% 
%At the phase boundary between CDW phase and $f$-PP phase, 
%we have 
%$\alpha_{CDW}=\alpha_{f-PP}=0$, indicating that the fermions (more precisely 
%the $f$-polarons) behave like free fermions at this point, due to
%the competition between 
%the acoustic coupling $g$ (giving an effective attractive interaction
%between fermions) and the effective forward scattering $G$ (repulsive after
%integrating out the $2k_f$ phonon field).
%Finally, we note that if we were to define the pairing
%operator by using bare fermions rather than $f$-polarons with optimal
%$\lambda_c$, we would have a region in the phase diagram in which
%none of the susceptibilities diverges (for the parameters used in
%Fig. \ref{phase_diag_alpha}(a) this regime extends between
%$0.55<g/v_f<0.68$). This is the so-called metallic phase regime 
%observed in Ref. \cite{martin}. However, as we will shown in Section
%\ref{impurity}, this regime is actually a superfluid phase because 
%a local impurity potential can be shown to be scaled to be irrelevant
%in the whole regime outside of the CDW phase. Therefore such superfluidity
%cannot be supported by the bare fermion pairing phase at least in the
%windows between CDW and BFP phase, where our $f$-polaron pairing mechanism
%can explain the superfluidity by its slowest algebraic decay..
%This supports of the polaron scenario proposed previously
%\cite{mathey} and in this paper.

In Fig. \ref{CDW_fPP} we show a schematic representation of a CDW and an
 $f$-PP phase.
In Fig. \ref{phase_diag_alpha}(b) we show a global phase diagram of a BFM by
considering the FP coupling ($g$) and the effective fermion-fermion interaction
($G$) as independent variables. 
The shading density indicates the strength of the phonon cloud
of the $f$-polaron pairing phase, $2\lambda_c$.
We find that for a fixed effective backward scattering between fermions,
i.e. $G=$constant, CDW phase, $f$-PP phase and phase separation regimes
become dominant successively when the long-wavelength fermion-phonon
coupling ($g$) is increased. 
%The superfluid $f$-PP regime becomes
%narrower when the effective (repulsive) interaction between fermions
%is increased. 
%The two types of coupling, $g$ and $G$, give rise to a competition 
% between CDW and $f$-PP: the FP coupling $g$, which creates an effective attractive
% interaction between the fermions, favors pairing, whereas
% $G$, which corresponds to an effective repulsion
% after integrating out the $2k_f$ phonons, supports CDW ordering.
%At the phase boundary we have $\alpha_{CDW}=\alpha_{f-PP}=0$, which
% indicates that the $f$-polarons behave like 
% free fermions at this point.
%%
%
%
%%
%
%
%
%-----------------
%
%
%
%
%

%The phase diagram in terms of 
% experimentally controlled parameters is shown 
%in Fig. \ref{phase_diagRP} and will be discussed in Section \ref{real_exp}. 
%
%
%
%
%
%
%
%%%%%%%%%%%%%%%%%
\subsection{Polaronic effects on bosons}
%
%Finally we note that polaronic effects can also be observed for the
%bosonic atoms as mentioned in the context of Eq. (\ref{b_polaron}).
The polaronic construction, demonstrated and discussed
 in the previous sections for the fermionic atoms,
 can also be done for the bosonic atoms, see Eq. (\ref{b_polaron}).
In Fig. \ref{phase_diag_alpha}(a), we also show the calculated scaling
exponent of the bare boson condensation field, $O_{BB}(x)=b(x)$, 
and that of the bosonic 
polaron ($b$-polaron) condensation field, $O_{b-P}(x)=\tilde{b}(x)$. 
In one dimensional systems,
the elementary excitations of the fermions are Luttinger bosons \cite{review},
which can provide the bosonic screening clouds around the bosonic atoms,
leading to a higher scaling exponent than the bare bosons as shown in Fig.
\ref{phase_diag_alpha}(a). However, in the fast boson limit ($v_b\gg v_f$)
 that we consider here,
the dressing of the bosonic atoms is very weak (i.e. $\alpha_{BB}\approx 
 \alpha_{b-P}$)
unless the system is close to the phase separation region. 
\subsection{Phase diagram in terms of experimental parameters}\label{PD_exp}
\begin{figure}
\includegraphics[width=8.5cm]{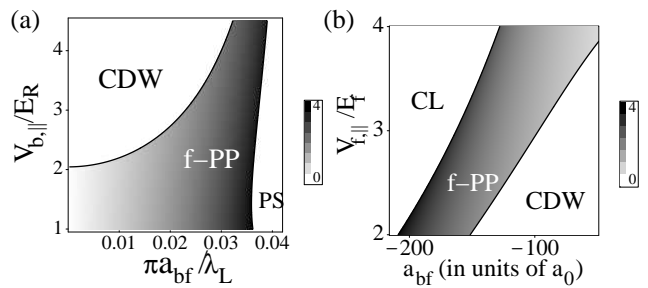}
\caption{
(a)
Phase diagram for a mixture of bosonic and spinless
fermionic atoms in a 1D optical lattice, for the
 first example discussed in Section \ref{PD_exp}. 
Shading in the $f$-PP phase describes the strength
of the bosonic screening cloud ($2 \lambda_c$, see Eq.(\ref{f_polaron})) around a pair of fermions. $\lambda_L$ and $E_R$ are the lattice period
and recoil energy, respectively.
Other parameters used for this figure are (see text for notations): 
$\nu_b=4$, $\nu_f=0.5$,
$V_{b,\perp}=V_{f,\perp}=20 E_R$, $V_{f,\|}=2 E_R$,  boson-boson 
scattering length $a_{bb}=0.01 \lambda_L$.
(b) 
Phase diagram for a mixture of bosonic and spinless
fermionic atoms in a 1D optical lattice, for the second example in
Section \ref{PD_exp}. 
Shading in the $f$-PP phase describes the absolute value of the strength
of the bosonic screening cloud ($|2 \lambda_c|$) around a pair of 
fermions. 
%$\lambda_L$ and $E_R$ are respectively the lattice period
%and recoil energy.
Other parameters used for this figure are (see text for notations): 
$\nu_b=3$, $\nu_f=0.2$,
$V_{f,\perp}/E_f=20$, $V_{b,\perp}/E_b= 1.65V_{f,\perp}/E_f$,  boson-boson 
scattering length $a_{bb}=100 a_0$, where $a_0$ is the Bohr radius. 
}
\label{phase_diagRP}
\end{figure}

In the last sections we derived 
 and discussed 
the $T=0$-phase diagram of BFMs.
% in terms of the 
%effective parameters $g$, $G$, $v_{f,b}$ and $K_b$. 
%We will now 
%discuss the experimental realization of these phases.
We will now give two plots of the same phase diagram
in terms of parameters that are experimentally more accessible.
% , i.e.
% by using the expressions (\ref{tbf}), (\ref{Ub}), and (\ref{Ubf}), and
% the Eqs. (\ref{vf}), (\ref{g}), (\ref{G}), (\ref{vb}) and (\ref{Kb}), 
% which relate the effective parameters to the lattice strengths, 
% scattering lengths and filling factors.
 In the first example (depicted in Fig. \ref{phase_diagRP} (a)), we
 assume that the lattice strengths $V_{f/b, \perp/ \|}$ can be
tuned independently, and 
%We choose fixed values for the two confining lattice strengths
% $V_{f/b, \|}$, and for the lattice strengths $V_{f,\||}$ of the
%laser that is parallel to the direction of the system, and couples to the fermionic
% atoms.
 we keep the densities $\nu_f$ and $\nu_b$, and the boson-boson
 scattering length $a_{bb}$ constant.
We now vary the boson-fermion scattering length $a_{bf}$ 
and the lattice strength $V_{b,\|}$ of the laser that is parallel to the system
direction and couples to the bosons.
This lattice strength mostly affects the bosonic tunneling $t_b$: For
 small $V_{b,\|}$, the tunneling amplitude is large, and therefore 
the phonons are fast.
% whereas f
 For large $V_{b,\|}$ the phonons are slow and the system develops CDW QLRO.
%The scattering length $a_{bf}$ appears linearly in $g$, and quadratic in $G$.
For large values of $a_{bf}$ the system experiences the 
 Wentzel-Bardeen instability, as shown in Fig. \ref{phase_diag_alpha} b).
%:
% One of the two eigenmode velocities, $v_A$, becomes imaginary. 
%As
%mentioned in Section \ref{Ham_diag}, this happens 
%for $\tilde{g}^2>\tilde{v}_f v_b$, which
% can be easily determined from the expression for $v_A$, Eq. (\ref{new_velocity}).
%The system in this regime will undergo phase separation, that is, the two
% liquids become immiscible and distribute themselves inhomogeneously in the
% trap.
%Within the stable regime
% we see the two phases that were discussed in the previous sections, CDW and $f$-PP.
%Slow bosons favor CDW over $f$-PP for two reason:
% Firstly, the smaller value of $\omega_{2k_f}$ reduces
% the energy cost of building up a 
% $2 k_f$ modulation in the bosonic superfluid.
% Secondly, as we show in Appendix \ref{A_polaron}, the polarization parameter 
%$\lambda_c$ is reduced, because slower phonons cannot screen the 
%fermions as effectively. This makes $f$-PP less favorable, because
% the effective attractive interaction between $f$-polarons is reduced.
%For larger values of $a_{bf}$ and for faster bosons, these considerations
% are reversed, and $f$-PP becomes dominant.
% In the diagram we also indicated the value of the parameter $2\lambda_c$, which
% describes the polarization of a Cooper pair. It increases as we approach the
% regime of phase separation (see also Appendix \ref{A_polaron}).

%---------
%
%

As a second 
%and an even more
% concrete 
 example 
 we will now discuss a ${^{87}}$Rb-${^{40}}$K mixture in an 
anisotropic optical lattice created by a set of Nd:YAG lasers.
In Fig. \ref{phase_diagRP} (b) we show a
phase diagram  for a mixture of bosons and spinless fermions
as a function of 
the scattering
length between bosons and fermions ($a_{bf}$ (in units of $a_0$))
%, which negative for most hyperfine states of interest of the two species, 
and the strength
of the longitudinal optical lattice for fermionic atoms ($V_{f,\|}/E_f$).
We assume that the lasers are far detuned from the transition energies, and that the optical lattices for the bosons and the fermions are created by the same set of lasers. In this set-up,
 %and in contrast to the first example, 
 the bosonic and fermionic lattice strengths are 
now constrained to the ratio: 
%, they are related by: 
$V_{b,\|/(\perp)}/E_b=1.65 V_{f,\|/(\perp)}/E_f$ for
 a laser wavelength of $830nm$ [\onlinecite{hmoritz}]. 
%(The other parameters for this example are given in the caption 
%of Fig. \ref{phase_diagRP}.)
We chose negative values for the boson-fermion scattering length, because 
for most combinations of hyperfine states the scattering length between ${^{87}}$Rb 
and ${^{40}}$K is negative.
%
%Independent tuning of the optical lattices for two
%species of atoms can be achieved even in the case when a single pair
%of lasers provides the standing beam in each direction.
%For example, the lattice strengths for bosons and fermions 
%in longitudinal direction is given by $V_{b, \|} \sim
%\Omega_b^2/\Delta$ and $V_{f, \|} \sim \Omega_f^2/(\delta - \Delta)$,
%where $\Delta$ is the detuning of the bosonic state, $\delta$ is the
%energy difference between the bosonic and the fermionic state, and
%$\Omega_{b/f}$ are the Rabi frequencies, which are propotional to the
%laser intensity. By controlling $\Delta$ and the laser intensity,
%$V_{b,\|}$ and $V_{f, \|}$ can be varied independently over a wide
%range.
%--------
%
%
%
%
Due to the different scaling of $v_b$ and $v_f$ with $V_{b,\|}/E_b$ and $V_{f,\|}/E_f$ the ratio of $v_b/v_f$ varies going
 along the vertical axis.  
 The lower portion of the diagram corresponds to slower bosons  
 (compared to the fermions) the upper part to faster bosons.
For relatively weak boson-fermion interactions and strong confinement  
%(i.e. 
%strong optical lattice for bosonic atoms) 
the system is
again 
 in the CDW phase, in which the densities of fermions and bosons have a $2 k_f$-modulation.
In the case of very strong boson-fermion interactions the system is 
again
unstable to
to a Wentzel-Bardeen instability, 
 which now leads to
collapse (CL) \cite{ho,instability},
 rather than phase separation.
 %because the boson-fermion interaction
  %is attractive. 
The two regimes are separated by an $f$-PP phase with an increasing value
 of the polarization parameter $\lambda_c$ for stronger interaction.
%While we carry out the detailed analysis for atoms in optical lattices, 
%qualitatively it also applies to the continuous
%case  \cite{bfm,sympathetic_cooling,general_1D}.  
%
%
%
%
%%%%%%%%%%%%%%%%%%%%%%%%%%%%%%%%%%%%
\section{Phase diagrams in other parameter regimes}
\label{other_parameter}
In the previous two sections we have considered BFMs 
in the limit of   
large bosonic filling ($\nu_b\gg\nu_f$) and fast bosons ($v_b\gg v_f$). 
We also assumed that both the fermionic filling and the 
bosonic filling are incommensurate to each other and to the lattice,
%Because of this assumption of incommensurability
so that 
 higher harmonics of the bosonization 
representation could be neglected.
We will now consider two cases in which 
%we 
%relax 
 these assumptions
 are relaxed in different ways: 
% and therefore have to include 
%additional terms in the Hamiltonian.
In section \ref{half_fill}, we consider a BFM  
with fermions at  half-filling.
%of fermionic half-filling. 
 At this particular fermionic filling, 
 phonon-induced 
Umklapp scattering can open a charge gap. 
%If it is relevant, a true phase transition occurs, resulting in a charge-gapped phase.
%In section \ref{comm_bf} we discuss the case when the 
%bosonic and the fermionic filling are commensurate to each other, 
%but incommensurate to the lattice. 
%Here, too, we need to include an additional nonlinear term in the Hamiltonian, that can be treated with an RG calculation.
%This case is most interesting when the bosons are close to the limit of hard-core bosons.  
%In subsection \ref{MI} we consider how the Mott insulator transition 
%of bosons at lattice-commensurate filling is influenced 
%by the presence of another superfluid, fermionic or bosonic.
In Sec. \ref{MI} we consider the case when the boson filling fraction 
is unity, so that the
 system can undergo a Mott insulator transition.
 A further discussion of commensurate mixtures is given in 
 \cite{commix}.
\begin{figure}
\includegraphics[width=4cm]{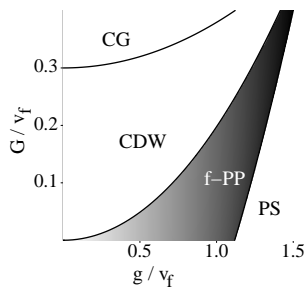}
\caption{Phase diagram of a BFM with spinless fermions at half-filling. All parameters are as in Fig.\ref{phase_diag_alpha} b. For large $G$ a charge gap (CG) appears due to the Umklapp scattering. }
\label{Half_Com}
\end{figure}
%
%
%
%
%
%
%
%
%%%%%%%%%%%%%%%%%%%%%%%%%
\subsection{Charge-gapped phase for half-filling of the fermions}
\label{half_fill}
In this section we still assume the system 
 to be in the regime of 
slow fermions ($v_f\ll v_b$) and large 
bosonic filling ($\nu_b\gg\nu_f$), but the 
fermionic filling is set to be at half-filling, 
i.e.  $\nu_f=1/2$. 
 In this situation, 
the derivation given in Eq. (\ref{eq:H_f^20}), leading to the 
 backscattering term Eq. (\ref{H_f^2}),
 has to include Umklapp scattering, 
which involves a momentum transfer of
%
%Because 
$4 k_f = 4 \pi \nu_f = 2 \pi$.
% is commensurate with the
%lattice constant,
%additional 
%terms, describing effective Umklapp 
%scattering, can also occur when integrating out the phonon field.
Similar to the derivation of Eq. (\ref{H_f^2}), 
 Umklapp scattering  can be obtained  to be:
%the elimination of the $2 k_f$-phonons also produces an Umklapp term, given by:
%
\begin{eqnarray}
H_{uk} &=& -\frac{1}{L} \frac{g_{2 k_f}^2}{\omega_{2 k_f}} \sum_{\alpha=\pm}
\sum_{k\sim 2\alpha  k_f} \sum_{p,q\sim \alpha k_f} f^\dagger_{p+k} f^{}_p
f^\dagger_{q+k} f^{}_{q}.
\label{eq:H_uk}
\end{eqnarray}
 In bosonized representation, this expression becomes:
%This can also be written as:
%
\bea
H_{uk} & = & - \frac{2 g_3}{(2\pi\alpha)^2} \int dx\cos\left[4 \theta_f(x)\right].
\eea
$g_3$ is the Umklapp parameter given
 by $g_{3}=\frac{g_{2 k_f}^2}{\omega_{2 k_f}}$, which 
 is equal to the backscattering parameter $G$. 
Since this is a nonlinear term that cannot be diagonalized, 
we use a scaling argument to study its effect.
 Such a renormalization flow argument at tree-level
 corresponds to a systematic expansion around $g_3 = 0$.  
The scaling dimension of this term is $4 K_\beta$, so 
 the flow equation for $g_3$ is given by:
\bea
\frac{d g_3}{dl} & = & (2 - 4 K_{\beta}) g_3
\eea
Therefore, the Umklapp term is relevant for $K_{\beta} < 1/2$. 
%(Note that CDW QLRO dominates already for $K_{\beta} < 1$). 
%In Fig. \ref{Half_Com} a), we show this additional phase transition. To fulfill $K_\beta <1/2$, we need at least $G>0.3v_f$. 
%This can be seen by noting that $\tilde{v}_f/v_A \cos^2\psi + \tilde{v}_f/v_a \sin^2\psi >1$ and rewriting $e^{2 \theta}=\Big(\frac{v_f - 2 G}{v_f + 2 G}\Big)^{1/2}<1/2$.
%In order to determine if this phase can be obtained in a real system, in which $G$ and $g$ are contrained, we need to consider the condition for phase separation as well.
%To avoid phase separation we need $4\tilde{g}^2<\tilde{v}_f v_b$. For the limit $v_f \ll v_b$, for which $G$ and $g$ can be simply related to the parameters of the original Hubbard model, as we did in section \ref{Bos_Ham}, it can be checked that this regime cannot be reached, since phase separation dominates.
When $g_3$ becomes relevant, the fermion excitations become gapped
due to the strong repulsion mediated by the $2 k_f$-phonons.
%This is analogous to  the Wigner crystal phase (WC).
 As  indicated in Fig. \ref{Half_Com}, 
 we find that 
%we indicate the region of the 
%phase diagram in which 
such a charge gap (CG) opens in the large $G$ regime.
\begin{figure}
\includegraphics[width=4cm]{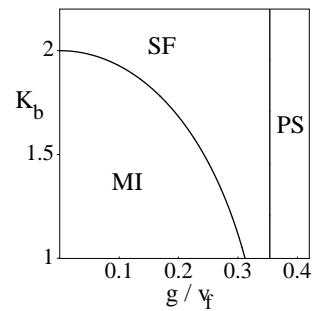}
\caption{Phase diagram of a BFM with unit bosonic filling. For small $K_b$ and small $g$ the bosons undergo a Mott insulator transition. 
In this diagram we only indicated the phases of the bosons. The fermions will be in an
 $f$-PP phase throughout the diagram.
}
\label{MI_melt}
\end{figure}
%
%
%
%
%
%
%
%%%%%%%%%%%%%%%
\subsection{Mott insulator transition for bosons at unit filling}
\label{MI}
In this section we determine how the Mott insulator (MI) 
transition in the bosonic sector is affected by the presence 
of the incommensurate fermionic liquid.
% Note, 
However, the results of this section 
equally apply to a boson-boson 
mixture because the dynamical properties of fermions
 and bosons in 1D are very similar. 
%We assume a mixture 
%with strongly repulsive bosons (corresponding to a small $K_b$), that 
%have a  filling that is commensurate to the lattice (here: $\nu_b=1$). 
The starting point for our discussion is not the single band 
Hubbard model (Eq. \ref{H_tot}), but a continuous 1D system, to which we add a 
weak $2 k_b$-periodic external potential:
\bea
H_{2 k_b} & = & \int dx V_{2 k_b} \cos 2 \theta_b.
\label{H2kb}
\eea
So the entire system is described by
\bea
H & = & H^{(0)}_b + H^{(0)}_f + H_{bf}^{(1)} + H_{2 k_b}
\eea
For simplicity we neglect the fermion backward scattering term.
%Note that there is no term as in Eq. (\ref{H_f^2}), since we have 
%comparable filling of bosons and fermions ($\nu_f \approx \nu_b$) and 
%typically also comparable velocities ($v_f \approx v_b$). 
 We again use a scaling argument to determine the
 effect of the non-linear term (\ref{H2kb}).
%The  term is of same form as described before,
 This term has a scaling dimension of $K_\delta$, and 
 its flow equation is therefore given by:
%
%
%
%Since the last term 
%is nonlinear, we again use a one-loop RG calculation, which gives us:
%
%
%
%
%
\bea
\frac{d V_{2 k_b}}{dl} & = & (2  -  K_\delta) V_{2 k_b}
\eea
The condition for the Mott insulator transition is therefore $K_\delta<2$. 
%which simplifies to $K_b<2$ at $g=0$.
% Since $v_b/V_A \sin^2 \psi + v_b/v_a \cos^2 \psi>1$, 
Since we always have $K_\delta>K_b$, the presence of 
the other atomic species tends to 'melt' the Mott insulator.
%, because $K_\delta >K_b$.
This can also be seen in Fig. \ref{MI_melt}: For $g=0$ we obtain a Mott insulator for $K_b<2$ and a superfluid for $K_b>2$. However, when we increase 
the fermion-phonon coupling
$g$ the Mott insulator regime shrinks and even 
vanishes entirely before phase separation is reached.
%Physically, 
 This is reasonable because
the coupling 
 to the second species of atoms 
 induces an attractive interaction, which tends to decrease
 the tendency of the bosons to form a Mott insulator at unit filling.
\section{Spinful system}
\label{spinful}
In this section we consider BFMs that have fermions with two internal
hyperfine states. 
%which we describe in a pseudospin language.
We assume the fermionic sector to be $SU(2)$ symmetric, that is, 
the two Fermi velocities and their filling
 fraction are equal, as well as their coupling to the bosonic atoms. 
 We again consider the limit of fast bosons, $v_b\gg v_f$.
\subsection{Hamiltonian}\label{Ham_S}
%
%
%%
%
%
%
%
%
%
%We will now derive the effective Hamiltonian
%of such a system, 
Analogous to the Hamiltonian of a BFM with 
spinless fermions, Eq. (\ref{H_eff}),
 we describe the two internal states with 
the fields $\Theta_{\uparrow, \downarrow}$ and $\Phi_{\uparrow, \downarrow}$.
In terms of these fields,
 the fermionic sector is described by a
 Hamiltonian of the form:
\bea\label{H_f^0_S}
H_f^0 & = & \int dx
 \Big\{
 \sum_{i=\uparrow,\downarrow} \frac{v_f}{2}
  \Big[ \frac{K_f}{\pi}
\left(\partial_x\Phi_{i}\right)^2+\frac{\pi}{K_f}\Pi_{i}^2 \Big]\nonumber\\
& & + U_{\uparrow \downarrow} \Pi_\uparrow\Pi_\downarrow
 +  \frac{2 U_{\uparrow \downarrow}}{(2\pi \alpha)^2}  
\cos(2\theta_{\uparrow}-2\theta_\downarrow) \Big\},
\eea
 where $\alpha\ra 0^{+}$ is the intrinsic
 cut-off of the bosonization representation.
The parameter $U_{\uparrow \downarrow}$ describes the 
 $s$-wave scattering between two spin states.
 %between
 %the two spin states, attractive or repulsive.
%Besides the term that couples the low-k part of the density fluctuations,
% $U_{\uparrow\downarrow}\int dx \Pi_\uparrow\Pi_\downarrow$,
% one also needs to include a non-linear term, 
%$\frac{2 U_{\uparrow \downarrow}}{(2\pi \alpha)^2} \int dx 
%\cos(2\theta_{\uparrow}-2\theta_\downarrow)$, 
 The last term is the backscattering term, which occurs
because 
 the two fermionic spin states have equal filling.
%Just as in the previous section, we will consider the RG flow of
% this non-linear term furtherdown.
%
%
%
%
The bosonic sector is 
of the same form as 
 before:
%for the spinless system:
%
%
%
\bea
H_b^0 & = & \frac{v_b}{2}\int dx
\left[\frac{K_b}{\pi}
\left(\partial_x\Phi_b\right)^2+\frac{\pi}{K_b}\Pi_b^2\right]
\eea
%
%
%
%
%The low-k part of the bosonic density fluctuations is coupled to the
%total density fluctuations of the fermions, 
Analogous to Eq. (\ref{H_bf^1}), the long wavelength
 density fluctuations
 of the fermions and the bosons
 are linearly coupled: 
\bea
H_{bf}^{(1)} & = & U_{bf} \int dx \Pi_b (\Pi_\uparrow +\Pi_\downarrow).
\eea
Finally, 
we also integrate out the $2k_f$-phonons of the bosonic superfluid, 
 leading to the following interaction between fermions:
%in the
%same way as for the spinless fermions. 
%In addition to terms of the form (\ref{H_f^2}) in each of the spin sectors,
% this also generates a non-linear term of the same form as in (\ref{H_f^0_S}):
%
%This term can be expressed in bosonized form as:
%
%
\bea
H_{f}^1 & = & \sum_{i=\uparrow,\downarrow} \frac{2G}{2\pi}\int dx
\left[\pi^2\Pi_i^2-(\partial_x\Phi_i)^2\right]\nonumber\\
& &  -\frac{8\pi G}{(2\pi\alpha)^2} \int
dx\cos\left(2\theta_\uparrow-2\theta_\downarrow\right)
\label{H_f^1_S}
\eea
The total low-energy Hamiltonian is then given by:
% the sum of these terms:
%
%
%
\bea
H & = & H_f^0 + H_b^0 + H_{bf}^{(1)} +  H_f^1,
\eea
%
%
%
%
%
%
%The low energy effective Hamiltonian can be written to be the sum of
%the following terms:
%
%\begin{eqnarray}
%H_{f, s}^0 &=& \frac{v_f}{2}\int_0^L dx\left[\frac{1}{\pi}
%\left(\partial_x\Phi_{f,s}(x)\right)^2+\pi\Pi_{f,s}(x)^2\right]
%\label{H_{f,s}^0}
%\\
%H_b^0 &=& \frac{v_b}{2}\int_0^L dx\left[
%\frac{K_b}{\pi}\left(\partial_x\Phi_b(x)\right)^2+
%\frac{\pi}{K_b}\Pi_b(x)^2\right]
%\label{H_b^0}
%\\
%H^1_{bf,s} &=& U_{bf}\int_0^L dx\, \Pi_b(x)\Pi_{f,s}(x)
%\label{H_{bf,s}^1}
%\\
%H_{f,s}^{1}&=&\frac{2G}{2\pi}\int_0^L
%dx\left[\pi^2\Pi_{f,s}(x)^2-(\partial_x\Phi_{f,s}(x))^2\right].
%\label{H_{f,s}^1}\\
%H_{f}^{2}&=& U_{\uparrow \downarrow} \int_0^L
%dx \Pi_{f,\uparrow}(x) \Pi_{f,\downarrow}(x) 
%\label{H_{f}^1}\\
%H_{\sigma, int} & =& \frac{2 g_{1\perp}}{(2\pi\alpha)^2} \int dx\cos\left[2(\Theta_{\uparrow}(x) - \Theta_{\downarrow})\right]
%\end{eqnarray}
%The spin symmetry of the system leads to a separation of the 
%Hamiltonian into a spin and a charge sector, 
%The total Hamiltonian 
 which 
can be separated into spin and charge 
sectors, 
$H=H_\rho +H_\sigma$, by introducing the following linear combinations:
\bea\label{spin_charge}
\Theta_{\rho/\sigma}  = \frac{1}{\sqrt{2}} (\Theta_{\uparrow} \pm \Theta_{\downarrow}), & & \Phi_{\rho/\sigma}  = \frac{1}{\sqrt{2}} (\Phi_{\uparrow} \pm \Phi_{\downarrow})
%\Theta_{\sigma}  = \frac{1}{\sqrt{2}} (\Theta_{\uparrow} - \Theta_{\downarrow}), & & \Phi_{\rho}  = \frac{1}{\sqrt{2}} (\Phi_{\uparrow} - \Phi_{\downarrow})
\eea
%
%
%
%
%
%
%The bosonic field couples only to the density ($\rho$) fields. 
%The charge part of the
%Hamiltonian 
 Note that $H_\rho$ 
coupled to the bosonic field 
is of the same form as a BFM with spinless fermions 
%$\theta_f \rightarrow\theta_\rho$. 
that we
discussed earlier and can be diagonalized 
by a rotation similar to
Eq. (\ref{Transformation_Engelsberg}). 
\begin{eqnarray}
\Pi_b=\delta_1\Pi_A+\delta_2\Pi_a, &&
\Phi_b=\epsilon_1\Phi_A+\epsilon_2\Phi_a,
\nonumber\\
\Pi_\rho= \beta_1\Pi_A+\beta_2\Pi_a, &&
\Phi_\rho= \gamma_1\Phi_A+\gamma_2\Phi_a,
\label{Transformation_Engelsberg2}
\end{eqnarray}
%With the parameters given by
%\begin{eqnarray}
%\begin{array}{ll}
%\beta_1 =  e^\theta \sqrt{\frac{\tilde{v}_\rho}{C_A}} \cos \psi  
%& \beta_2  =  e^\theta \sqrt{\frac{\tilde{v}_\rho}{C_a}} \sin \psi\\
%\gamma_1  =  e^{-\theta} \sqrt{\frac{C_A}{\tilde{v}_\rho}} \cos \psi 
%& \gamma_2   = e^{-\theta} \sqrt{\frac{C_a}{\tilde{v}_\rho}} \sin \psi\\
%\delta_1 =  - e^\phi\sqrt{\frac{v_b}{C_A}} \sin \psi  
%& \delta_2  =  e^\phi\sqrt{\frac{v_b}{C_a}} \cos \psi\\
%\epsilon_1  =  - e^{-\phi}\sqrt{\frac{C_A}{v_b}} \sin \psi 
%& \epsilon_2   =  e^{-\phi}\sqrt{\frac{C_a}{v_b}} \cos \psi.
%\end{array}
%\label{coefficients2}
%\end{eqnarray}
%
%Here $e^\theta = 
%((v_\rho - 2 G)/(v_\rho + 2 G))^{1/4}$,
%$e^{\phi}=\sqrt{K_b}$, and
%$\tan 2\psi = 4 \tilde{g} (v_b \tilde{v}_\rho)^{1/2}/(v_b^2 - \tilde{v}_\rho^2)$.
The diagonalized Hamiltonian of the charge sector has the same form as (\ref{new_hamiltonian}):
\begin{eqnarray}
H_\rho&=&\frac{1}{2}\sum_{j=a,A}v_j\int dx\left[\pi\Pi_j(x)^2
+\frac{1}{\pi}\left(\partial_x\Phi_j(x)\right)^2\right]
\label{new_hamiltonian_spinful}
\end{eqnarray}
and, similarly, the eigenmode velocities, $v_A$ and $v_a$, are 
of the same form as (\ref{new_velocity}):
\begin{eqnarray}
v_{a/A}^2 & = & \frac{1}{2} (v_b^2 + v_\rho^2) \pm \frac{1}{2} 
\sqrt{(v_b^2 - v_\rho^2)^2 +  16 \tilde{g}_\rho^2  v_b v_\rho},
\label{new_velocity_spinful}
\end{eqnarray}
where we defined $v_\rho\equiv(\tilde{v}_\rho^2-4G_\rho^2)^{1/2}$ 
and $\tilde{g}_\rho\equiv
g_\rho\,e^\theta$, with $e^\theta = 
((\tilde{v}_\rho - 2 G_\rho)/(\tilde{v}_\rho + 2 G_\rho))^{1/4}$, $g_\rho$ by $g_\rho = \sqrt{2} g$. $\tilde{v}_\rho$ is given by $\tilde{v}_\rho = v_f + G$, $G_\rho = G + G_{\uparrow \downarrow}/2$. $G_{\ua\da}$ is given by $G_{\ua\da}=U_{\ua\da}/2\pi$.
All of these parameters and expressions can be immediately obtained by 
transferring the results from the spinless BFM case.
% after applying the
% the transformation (\ref{spin_charge}).
%
%
%

%
%
%
The quadratic part of the spin part of the
Hamiltonian can be diagonalized 
%--- except for the back-scattering term --- 
by using 
\begin{eqnarray}
\Pi_{\sigma}= \sqrt{K_\sigma} \tilde{\Pi}_\sigma, &&
\Phi_{\sigma}= 1/\sqrt{K_\sigma} \tilde{\Phi}_\sigma,
\end{eqnarray}
where  $\sqrt{K_\sigma} \equiv e^{\phi_\sigma}$ and 
$\tanh 2\phi_{\sigma}=-2G_\sigma/\tilde{v}_\sigma$, with 
$\tilde{v}_{\sigma} = v_f - G_{\uparrow \downarrow}$,  
$G_{\sigma} = G - G_{\uparrow \downarrow}/2$ and 
$v_\sigma=\sqrt{\tilde{v}^2_\sigma - 4 G_\sigma^2}$. 
The transformed 
spin sector then becomes a sine-Gordon model, 
\bea
H_\sigma & = &
\frac{1}{2}v_\sigma\int dx\left[\pi\Pi_\sigma(x)^2
+\frac{1}{\pi}\left(\partial_x\Phi_\sigma(x)\right)^2\right]\nn\\
& & +\frac{2
g_{1\perp}}{(2\pi\alpha)^2} \int
dx\cos\left(\sqrt{8K_\sigma}\Theta_\sigma\right).
\eea
%
%
%
%
%
%
%where we have (for simplicity) renamed $\tilde{\Theta}_\sigma$ by $\Theta_{\sigma}$.
%$v_\sigma$ is the spin velocity, $K_\sigma$ is the spin Luttinger
%exponent, and 
Here, $g_{1\perp}=U_{\uparrow\downarrow} -4\pi G$ is the
effective backward scattering amplitude for fermions.
%, that has
%contributions from fermion-fermion interactions 
%($U_{\uparrow\downarrow}$) and from integrating
%out $2k_f$ phonons ($G$).  
%
%%
%
%
%
%
%
%
\subsection{Renormalization flow}\label{RG_S}
\begin{figure}
\includegraphics[width=5cm]{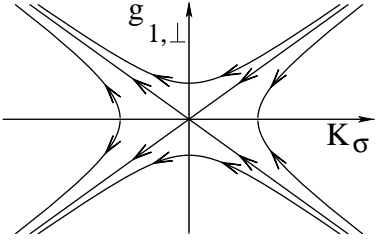}
\caption{
Linearized renormalization flow of the parameters $g_{1,\perp}$ and $K_\sigma$
in the vicinity of the SU(2) symmetric 
fixed point at $g_{1,\perp}=0$ and $K_\sigma=1$.
%The SU(2) symmetry of the 
Systems with SU(2) symmetry are forced
to be on the separatrix from the upper right 
quadrant to the lower left quadrant.
If the initial parameters of such a system correspond to a point in
 the upper right quadrant the system flows 
to the non-interacting, SU(2) symmetric
 fixed point at the origin. For initial parameters in the lower
 left quadrant 
the system flows towards a strongly 
interacting fixed point and opens a spin gap.
}
\label{SG_flow}
\end{figure}
In this section,  
%we discuss the behavior of the spin sector of the system.
%Since its Hamiltonian contains a non-linear term due the commensurability
% of the two fermionic liquids,
 we deal with the spin Hamiltonian by performing an RG calculation, 
 similar to the calculation in section \ref{other_parameter}.
As discussed in \cite{review}, and as is
well-established in the literature,
 the RG flow at one-loop is given by:
\bea
\frac{d K_\sigma}{d l} & = & - \frac{K_\sigma^2}{2} \Big(\frac{g_{1, \perp}}{\pi v_\sigma}\Big)^2\\
\frac{d g_{1, \perp}}{d l} & = & (2 - 2 K_\sigma) g_{1, \perp}
\eea
 These equations are a systematic expansion
 around the non-interacting fixed point
 at $g_{1,\perp}=0$ and $K_\sigma=1$. 
In Fig. \ref{SG_flow}, we show the linearized flow 
generated by these equations, in the vicinity of 
%the non-interacting
 this fixed point.
% at $g_{1,\perp}=0$ and $K_\sigma=1$.
 For $K_\sigma>1$,  
%(corresponding to positive $g_{1,\perp}$) 
the parameters of the system
 flow towards the fixed point $(g_{1,\perp}=0, K_\sigma = 1)$, 
whereas for $K_\sigma<1$,
% (corresponding to negative $g_{1,\perp}$) 
they flow towards the strong coupling fixed 
point $(g_{1,\perp} \rightarrow \pm \infty, K_{\sigma} =0)$.
The nature of spin excitations of 
the ground state follows from the well known properties of the
sine-Gordon model for SU(2) symmetry: 
When the initial $g_{1\perp}$ is positive the system has
gapless spin excitations,  i.e. $g_{1\perp} \ra 0$ and $K_\sigma\ra 1$,
%($g_{1\perp}$ is irrelevant in the RG
%sense), 
and when the initial $g_{1\perp}$ is negative the system has a spin gap, 
 i.e. $g_{1\perp}\ra -\infty$ and $K_\sigma \ra 0$.
%($g_{1\perp}$ tends to a strong coupling fixed point under the
% RG flow). 

%
%The effect of this RG flow on the quasi phases can be seen from the scaling exponents (see Table II).
%
%
%
%
%We will determine the phase diagram by studying the scaling of different order parameters, as we did for the case of BFM with spinless fermions.
%
%
%
%
\subsection{Scaling exponents}\label{scaling_S}
%
%
%
%
%
%
%
%\begin{figure}
%\includegraphics[width=8cm]{spinfulalphas2.eps}
%\caption{
%Scaling exponents of different order parameters. In both figures $v_b/v_f=5$, $K_b=10$ and $U_{\uparrow\downarrow}/v_f=0.8\pi$. 
%In (a) $G/v_f=0.1$, in (b) $G/v_f=0.25$.
%Due to the opening of the spin gap the CDW and the SPP exponent have been lifted up by $1$, because $K_\sigma$ has been sent from $1$ to $0$, whereas the SDW and the TPP exponents have been sent to $- \infty$.
%}
%\label{spinalphas}
%\end{figure}
%
%
%
%
%
Now we will determine the scaling exponents
 of the order parameters that can show QLRO in the phase diagram.
%In the following section we
 %will use these scaling exponents to determine
%the phase diagram, as we did for BFMs with spinless fermions.
%
%
%
%
The order parameters of a standard SU(2) symmetric
 LL of fermions
%that appear in the phase diagram 
%of BFM with spinful fermions 
are the charge density wave (CDW), spin density wave (SDW), singlet and triplet
 pairing (of bare fermions), and Wigner crystal (WC) operator.
When such a system is coupled
 to a bosonic superfluid
 we again need to replace the pairing of bare fermions 
by pairing 
 of polarons (analogous to Section \ref{scaling}, see also Appendix \ref{A_scaling}).
% resulting in the introduction 
%  of singlet polaron pairing (SPP) and triplet polaron pairing (TPP).
First we consider the $2 k_f$-mode of the CDW operator ($O_{CDW}=\sum_s f_{s,L}^\dagger f_{s,R}$) and of the SDW operator ($O_{SDW}=\sum_{s,s'} f_{s,L}^\dagger \hat{\sigma}_{s,s'} 
f_{s',R}^{}$). In bosonized form these correspond to $O_{CDW, 2 k_f} \sim \exp[\sqrt{2} i \Theta_\rho + \sqrt{2} i \Theta_\sigma]$ and $O_{SDW, 2 k_f} \sim \exp[\sqrt{2} i \Theta_\rho + \sqrt{2} i \Phi_\sigma]$. Their scaling exponents are given by
\bea
\alpha_{CDW} & = & 2 - K_\beta - K_\sigma\\
\alpha_{SDW} & = & 2 - K_\beta - K_\sigma^{-1}
\eea
%
%Due to the larger number of degrees of freedom more types of 
%ordering are possible: in addition to the CDW phase, we have a 
%spin density wave (SDW), where the spin order parameter at wave vector $2 k_f$ 
%($O_{SDW}=\sum_{s,s'} f_{s,L}^\dagger \hat{\sigma}_{s,s'} 
%f_{s',R}^{}$ with $\hat{\sigma}_{\alpha\beta}$ being the Pauli matrix) 
%developes quasi-long-range order.
For the polaron pairing phase, both singlet (SPP,
$O_{SPP}=\tilde{f}_{\uparrow,L} \tilde{f}_{\downarrow,R}^{}
-\tilde{f}_{\downarrow,L} \tilde{f}_{\uparrow,R}^{}$) and triplet 
(TPP, $O_{TPP}=\tilde{f}_{\uparrow,L}
\tilde{f}_{\uparrow,R}$)
pairing needs to be considered.
 In bosonized form, 
%The $k=0$-mode of 
these operators are 
given by $O_{SPP,k=0}\sim\exp[-2 i \lambda_c \Phi_b + \sqrt{2} i \Phi_\rho + \sqrt{2} i \Theta_\sigma]$ and $O_{TPP, k=0}\sim \exp[-2 i \lambda_c \Phi_b + \sqrt{2} i \Phi_\rho + \sqrt{2} i \Phi_\sigma]$, 
 with scaling exponents:
%. These operators scale with the exponents
%
\bea
\alpha_{SPP} & = &
% 2 - [K_\gamma^{-1} + 2 \lambda_c^2 K_\epsilon^{-1} - \sqrt{8} \lambda_c K_{\gamma \epsilon}^{-1} +K_\sigma]\\
%2 - [K_\gamma^{-1}  - K_\epsilon/K_{\gamma \epsilon}^{2} +K_\sigma]
%\label{alpha_SPP_0}\\
%& = & 
2 - [K_\beta^{-1} + K_\sigma]
\label{alpha_SPP}\\
\alpha_{TPP} & = & 
%2 - [K_\gamma^{-1} + 2 \lambda_c^2 K_\epsilon^{-1} - \sqrt{8} \lambda_c K_{\gamma \epsilon}^{-1} +K_\sigma^{-1}]
%2 - [K_\gamma^{-1}  - K_\epsilon/K_{\gamma \epsilon}^{2} +K_\sigma^{-1}]
%\label{alpha_TPP_0}\\
%& = & 
2 - [K_\beta^{-1} + K_\sigma^{-1}]
\label{alpha_TPP}
\eea
%
%Some of these quantum phases can coexist in certain
%regimes of the phase diagram (see Fig. \ref {phase_diag_spin}).
These expressions 
%(\ref{alpha_SPP_0}) and (\ref{alpha_TPP_0}) 
were obtained by using $\lambda_c = K_\epsilon/(\sqrt{2} K_{\gamma \epsilon})$,
 and 
%For (\ref{alpha_SPP}) and (\ref{alpha_TPP}) we used 
the identity $K_\beta^{-1}=K_\gamma^{-1} - K_\epsilon K_{\gamma \epsilon}^{-2}$, as we did for spinless BFMs.
Again, we can see that these many-body order parameters
have scaling parameters controlled by two parameters only,
$K_\beta$ and $K_\sigma$,
indicating that the spinful BFM system can also be understood as a LL of polarons.
% if $K_\beta$ is understood as the LL parameter
% of polarons.

%
%
%
%
%
We
 now discuss how these scaling exponents behave in different parameter regimes.
We consider the case in which the spin sector flows towards the Gaussian fixed point, $K_\sigma=1$.
%In Fig. \ref{spinalphas} (a), we see the scaling exponents of SDW and CDW as well as TPP and SPP, for $G/v_f=0.1$.
%For this value we have a gapless spin phase, because $g_{1,\perp}$ is positive. 
%(This value of $G$ corresponds to a horizontal line in Fig. \ref{phase_diag_spin} (b).). 
%According to the RG flow we have $K_\sigma = K_\sigma^{-1}=1$ in this regime. 
Therefore, SDW and CDW as well as TPP and SPP are degenerate  at the level of algebraic exponents. However, by taking into account logarithmic corrections \cite{review}, SDW and TPP turn out to have slower decaying correlation functions and are therefore dominant.
As a result, in the gapless phase there are two regimes of QLRO: SDW (CDW) and TPP (SPP), where the brackets refer to subdominant ordering.
%
%In Fig. \ref{spinalphas} (b), at $G/v_f=0.25$, 
For large $G$, the system is in the 
spin-gapped phase, because $g_{1, \perp}$ is negative now, and therefore becomes
 relevant.
 The system then 
flows to the strong coupling fixed point and $K_\sigma \rightarrow 0$, according to the RG flow. 
%Therefore CDW and SPP are lifted up by one, whereas SDW and TPP have been sent to $-\infty$.
%As a result, we find that  
%there are now four regimes: For small $g$, there is only CDW ordering, for large $g$, close to phase separation, there is only SPP ordering. In between, these types of ordering overlap: 
%there is one regime in which CDW is dominant and SPP subdominant, and another one with the opposite situation.
%These scaling exponents are used to obtain phase diagrams.
%
%
%
%
%
%
\subsection{Phase diagram}\label{PD_S}
\begin{figure}
\includegraphics[width=8cm]{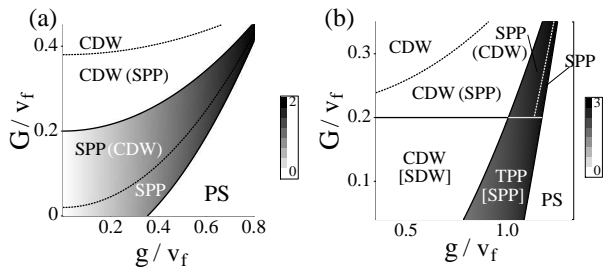}
\caption{
Phase diagrams for a mixture of bosonic and $S=1/2$ fermionic 
atoms with $v_b/v_f=5$ and $K_b=10$. 
In (a) $U_{\uparrow\downarrow}/v_f=-0.8\pi$
and in (b) $U_{\uparrow\downarrow}/v_f=0.8\pi$.
%
%At the top portion of (b) and everywhere in (a) we have
%$g_{1\perp}<0$ ($K_{\sigma}<1$), and $g_{1\perp}<0$ 
%($K_{\sigma}>1$) at the bottom of (b). 
%For larger $G$ and $U_{\uparrow\downarrow}$ one can 
%also have a Wigner crystal phase (not shown here). 
Parentheses ($\dots$) indicate subdominant phases.
%Phase diagrams for a mixture of bosonic and $S=1/2$ fermionic 
%atoms. Shading in the paired regions describes the strength
%of bosonic screening clouds around fermionic
%polarons. In both figures $v_b/v_f=5$ and $K_b=10$. 
%In (a) $U_{\uparrow\downarrow}/v_f=-0.8\pi$, in (b) $U_{\uparrow\downarrow}/v_f=0.8\pi$.
%At the bottom of (b) and everywhere in (a) we have
%$g_{1\perp}>0$ (see text), and $g_{1\perp}<0$ at the top of (b). 
%Note that 
%parentheses ($\dots$) indicate subdominant phases, while square 
%brackets [$\dots$] indicate degenerate phases. 
}
\label{phase_diag_spin}
\end{figure}
%
%
%
%
%The scaling exponents derived in the previous section
% can now be used to construct the phase diagram.
%
%
%
%
In Fig. \ref{phase_diag_spin} we show two examples for different values of $U_{\uparrow, \downarrow}$.
In Fig. \ref{phase_diag_spin} (a) we show the phase diagram for $U_{\uparrow, \downarrow}/v_f =-0.8\pi$,
 and  in Fig. \ref{phase_diag_spin} (b) for $U_{\uparrow, \downarrow}/v_f =0.8\pi$.
%, which is the value that has been used for Fig. \ref{spinalphas}.
%As for the case of spinless fermions, for large $g$ the system becomes unstable to phase separation (PS). For small values of $g$, density-wave orderings dominate, either SDW order (in the gapless phase)  or CDW (in the presence of a spin gap). In between these two regimes, there is a pairing regime, either TPP or SPP.
As described above, the system develops a spin gap for $g_{1 \perp}<0$. The phase transition occurs at $4 \pi G = U_{\uparrow, \downarrow}$. In Fig. \ref{phase_diag_spin} b), this corresponds to $G/v_f=0.2$,  
%The regime above this value correspond to a scenario similar to Fig. \ref{spinalphas} (b), the regime below to Fig. \ref{spinalphas} (a).
 in Fig. \ref{phase_diag_spin} (a), the entire phase 
diagram is in the gapped phase, because $U_{\uparrow,\downarrow}<0$, 
and $G$ can only be positive. 
%So for the entire diagram, the scaling exponents behave similar to Fig. \ref{spinalphas} (b).
% In
%order to describe possible ground states of the system we calculated
%the low temperature behavior of susceptibilities for the following
%order parameters: $2k_f$ spin density wave ($\chi_{SDW}$), $2k_f$
%charge density wave ($\chi_{CDW}$), $4k_f$ (Wigner crystal) charge
%density wave ($\chi_{WC}$), singlet polaron pairing ($\chi_{SPP}$), and
%triplet polaron pairing ($\chi_{TPP}$). Depending on the parameters we
%find the following regimes: 1) Equally divergent $\chi_{CDW}$ and
%$\chi_{SDW}$.  Degenerate CDW and SDW phases; 2) Equally divergent
%$\chi_{SPP}$ and $\chi_{TPP}$.  Degenerate SPP and TPP phases; 3)
%Divergent $\chi_{SPP}$. State with singlet pairing of polarons; 4)
%Both $\chi_{CDW}$ and $\chi_{SPP}$ diverge, but unequally. This probably
%corresponds to a phase that has both a CDW order and singlet polaron
%pairing; 5) Divergent $\chi_{CDW}$. CDW phase; 6) Divergent
%$\chi_{WC}$. Wigner crystal phase. Phase diagram 
%in Fig. \ref{phase_diag_spin} again 
%shows remarkable similarity to a phase diagram for interacting
%electrons \cite{solyom}.
%------
%
%
%
%%
%
%
%
%
%
\subsection{Wigner crystal}\label{WC_S}
Apart from these phases the system can also show Wigner crystal (WC) ordering.
%
%, as shown in
%Fig. \ref{wigner}.
Within the WC phase, the fermionic atoms crystallize in an alternating pattern of spin-up and spin-down atoms, giving rise to a $4 k_f$-density modulation.
The Wigner crystal order parameter is given by $O_{WC}=\sum_s f^{\dagger}_{R,s} f^{\dagger}_{R, -s} f_{L,-s} f_{L,s}
%The $4 k_f$-mode of this operator is described by $O_{WC, 4 k_f}
\sim\exp[\sqrt{8} i \Theta_\rho]$ and has the scaling exponent
\bea
\alpha_{WC} & = & 2 - 4 K_\beta
\eea
Within the gapless phase ($K_\sigma=1$), the Wigner crystal phase can become dominant for $K_\beta < 1/3$, which can be achieved for large values of $G$. 
%In order to avoid the spin-gapped phase we also need a large value of $U_{\uparrow\downarrow}$, so that $g_{1,\perp}$ is still positive.
%
%\begin{figure}
%\includegraphics[width=9cm]{wigner.eps}
%\caption{Emergence of the Wigner crystal phase. Parameters for (b): $v_b/v_f=5$, $K_b=10$ and $U_{\uparrow\downarrow}/v_f=1.6\pi$. The Wigner crystal phase (WC) has CDW and SDW as subdominant phases. For (a), we chose $G/v_f = 0.38$, which corresponds to a horizontal line below the spin gap formation at $G/v_f=0.4$ in (b). 
%}
%\label{wigner}
%\end{figure}
%
%
%
%
%
%
%
%
%
%
%
%
%%%%%%%%%%%%%%%%%%%%%%%%%%%
\section{Experimental Issues}
\label{discussion}
%%%%%%%
%
%
%
%
%%%%%%%%%%%%%%%%%%%%%%%%%%%%%%%
%\subsection{Experimental Realization}
%\label{real_exp}
%
%
%
%
In this paper we have calculated 
 the phase diagrams
 of different types of BFMs in various parameter regimes.
 In this section we will discuss issues concerning 
  the experimental realization of our systems.

Throughout the paper we concentrated on
 infinite systems at 
zero temperature, 
$T=0$. The realization of 1D Luttinger liquids of ultracold atoms will of course be in a finite lattice (around 100 sites) and at finite temperature.
At finite temperature the correlation functions become 
 approximately 
$\langle O(x) O(0)
\rangle \sim (\xi \sinh(x/\xi))^{\alpha-2}$, 
 where $\xi\sim v_f/k_B T$
 is the thermal correlation length. 
%The system acquires a 
% thermal correlation lengths for $O_{CDW}$ and $O_{f-PP}$, which
% are approximately given by $\xi \sim \beta v_f$, 
% where $\beta$ is the inverse
% temperature. 
For a finite system of length $L=N \lambda_L$ 
($N$ being the number of lattice sites), the $T=0$-properties 
of the system are visible for $\xi \sim L$. 
This corresponds to a temperature regime of 
$T \sim T_f/N$, with $T_f$ being the Fermi temperature. 
So for a system with $N=100$, $T\sim 0.01 T_f$ is required.

%\subsection{Detection schemes}

Another issue is the experimental signature of the different phases. 
Here we present
 several approaches that can be used to detect the quantum phases discussed above. 
%One approach, t
To detect the CDW phase, one can
 perform
a standard time-of-flight (TOF) measurement.
In the CDW phase the fermion density modulation will
induce a $2k_f$ density wave
in the boson field in addition to the zero momentum 
condensation so that the CDW phase can be observed as interference peaks at momentum
$k=2k_f$ in a TOF measurement for bosons, 
 whereas the $f$-PP phase will show a featureless
 superfluid signature. 
 We note that 
%a homogeneous system in 1D can only 
%develop 
%quasi-long-range order. However, 
the boundary of an atomic trap
 and other inhomogeneities
in a realistic experiment can pin the CDW phase and generate a true
density modulation.
%%
%
%Therefore in principle the phase transition from the CDW phase 
%toward the pairing phase can be observed by comparing the
%data of bosons and fermions obtained in the TOF measurement,
%which can be easily acheived in the present experiments \cite{endnote2}. 
%
%Another direction is to 
One may also use a laser stirring 
experiment \cite{stirring}  to probe 
the phase boundary between the insulating (pinned by trap potential) 
CDW and the superfluid $f$-PP phase: a laser beam
is focused at the center of the cloud to
 create a local potential, 
 and is then moved oscillatory in the condensate.
%  to measure the response of the BFM.
If the system is in the pairing phase, the
laser beam can be moved through the system without
dissipation, which would manifest itself as heating,
  if only its velocity is slower than some critical
value \cite{stirring}. At the $f$-PP/CDW phase
boundary this critical velocity goes to zero, reflecting a transition
to the insulating (CDW) state. This scenario follows from 
the RG analysis of a single impurity potential \cite{kane}, 
 as described in Appendix
 \ref{impurity}.
%Finally, a way 
To probe the boundary of phase separation  
 one can 
measure the dipolar collective oscillations of the system,
generated by a sudden
displacement of the harmonic trap potential with respect to 
the lattice potential
\cite{collectivemodes}. 
When the system is near the PS boundary, fermion-boson interaction 
will reduce the frequency of the dipolar mode 
 essentially to zero, because the two atomic species
 become immiscible.

As the most promising approach
 we propose to study the noise correlations of TOF measurements, as
 discussed in [\onlinecite{noise_corr}].
This type of measurement treats particle-particle correlations 
(i.e. pairing fluctuations) and particle-hole correlations
 (i.e. CDW and SDW fluctuations) on equal footing, and therefore
 reflects the formal duality of these two types of phases accurately.
 Furthermore, this approach seems to be well-suited for the study of 
 Luttinger liquids, because it gives signatures of the various fluctuations
 in the system, 
%among which one or several dominate to create QLRO.
 that dominate in different regimes of the phase diagram.

%%%%%%%%%%%%%%%%%%%%%%%
\section{Conclusion}\label{conclusion}
In summary, we used bosonization to
investigate the quantum phase diagrams of 1D BFMs.
%mixtures of bosonic and fermionic atoms involving spinless
%and $S=1/2$ fermions in various regimes. 
Interactions between atoms can lead to 
interesting phenomena like spin and charge density waves, singlet and
triplet pairing of atoms.  We introduced polarons, i.e. atoms of one
species surrounded by screening clouds of other species, and argued
that the rich phase diagrams of BFMs can
 be  naturally 
interpreted as Luttinger liquid phase diagrams of such polarons.
We also considered several commensurate 
filling cases 
%of these systems,
 %and determined the phase diagram, which
% now includes 
 and obtained 
 gapped phases in some parameter regimes.
%in which regimes they can lead to a gapped phase.
 We discussed several techniques for
probing our results experimentally. 
%like paired and insulating phases
%of fermions.

We thank E. Demler, M. Lukin, H. Moritz 
and B.I. Halperin for useful discussions.
%This work was supported by the NSF (grants DMR-01328074, PHY-0134776),
%the Sloan and the Packard Foundations, and by Harvard-MIT CUA.
%
%
%
%
%
%
%%%%%%%%%%%%%%%%%%%%%%%%%%%%%%
\appendix
%%%%
%
%
%
%
%
%
\section{Scaling Exponents}\label{A_scaling}
In this section we will derive the scaling exponents of 
 a number of operators that 
 were considered in the search for the phase diagram.
%are of importance for the study of BFM.
In section \ref{A_scaling_SL}, we discuss BFMs with spinless fermions, 
 and in
 section \ref{A_scaling_S} BFMs with spinful fermions.
\subsection{Spinless Fermions}\label{A_scaling_SL}
%
%
%
%
%
% For that purpose 
%We consider a
%general operator (cp. \cite{cazalilla}), which can then be 
% specialized to the operators appearing in the discussion. 
%
%
%
%
For the case of spinless fermions, a broad 
 class of operators $O(x)$  (cp. \cite{cazalilla}) can be written in 
terms of the Luttinger fields as  
\bea
O(x) & \sim & \exp[\sum_{j = b,f} (i \lambda_{j,1} \Theta_j + i\lambda_{j,2} \Phi_j)]
\eea
or a sum of products of this type. $\lambda_{j,1/2}$ are 
arbitrary real parameters.
To derive the correlation function of such an operator we first write $\Theta_{b/f}$
 and $\Phi_{b/f}$ in terms of the eigenfields $\Theta_{A/a}$ and $\Phi_{A/a}$, as
given by the transformation (\ref{Transformation_Engelsberg}).
%Since the Hamiltonian, written in these fields, is just a sum of Gaussian 
%models, each of these fields has a correlation function that behaves logarithmically
% for large distances, 
%i.e. 
%$\lav(\theta_{A/a}(x)-\theta_{A/a}(0))^2\rav\sim \log |x|$
%and 
%$\lav(\Phi_{A/a}(x)-\Phi_{A/a}(0))^2\rav\sim \log |x|$.
%With these results,
 The correlation function of the
above given general operator,
$C_O (x) = \lav O(x) O(0)\rav$, then behaves for large distances as :
\bea
C_O(x) & \sim & |x|^{-\tilde{\alpha}} 
\cos((\lambda_{f,1}\nu_f + \lambda_{b,1}\nu_b)\pi x),
\eea
 with the scaling exponent  
$\tilde{\alpha}$  given by
\bea
\tilde{\alpha} & = & \frac{1}{2} (\lambda_{b,1}^2 K_\delta + \lambda_{f,1}^2 K_\beta + 2 \lambda_{b,1} \lambda_{f,1} K_{\beta \delta}\nonumber\\
& & +   \lambda_{b,2}^2 K_\epsilon^{-1} + \lambda_{f,2}^2 K_\gamma^{-1} + 2 \lambda_{b,2} \lambda_{f,2} K_{\gamma \epsilon}^{-1})
\eea
From this general expression we can deduce the scaling exponents discussed 
in this paper, by specializing the operator $O(x)$ to 
 various cases. 

\subsection{Spinful fermions}\label{A_scaling_S}
For the case of spinful fermions, we consider 
an operator $O(x)$ given by:
\bea
O(x) & \sim & \exp[\sum_{j = b,\rho,\sigma} (i \lambda_{j,1} \Theta_j + i\lambda_{j,2} \Phi_j]
\eea
%
%
%
%
%
%or a sum of products of this type. 
To determine the correlation function of this operator, we 
 proceed as in the previous section. First, we
 use the linear transformation (\ref{Transformation_Engelsberg2}), to 
 write $\theta_{\rho/b}$ in terms of the fields $\theta_{A/a}$.
Since the density/boson--sector of the Hamiltonian is just a sum 
 of Gaussian models in terms of these fields, the scaling exponents
 can be determined as in \ref{A_scaling_SL}. 
 For the spin sector we use the RG results discussed in Section \ref{RG_S}: If
 the non-linear term ir irrelevant the system is described
 by a Gaussian model with $K_\sigma=1$. If the non-linear term
 is relevant we have $K_\sigma\rightarrow 0$, i.e. any overlap of the operator
 with the phase field $\Phi_\sigma$ will make the correlation function 
 short-ranged, whereas the field $\theta_\sigma$ acquires ordering.

The correlation function 
 of the operator $O(x)$,
$C_O (x) = \lav O(x) O(0)\rav$, can be determined to be:
\bea
C_O(x) & = & |x|^{-\tilde{\alpha}} \cos((\lambda_{f,1}\nu_f + \lambda_{b,1}\nu_b)\pi x)
\eea
The scaling exponent $\tilde{\alpha}$ is given by
\bea
\tilde{\alpha} & = & \frac{1}{2} (\lambda_{b,1}^2 K_\delta + \lambda_{\rho,1}^2 K_\beta + 2 \lambda_{b,1} \lambda_{\rho,1} K_{\beta \delta}\nonumber\\
& & +   \lambda_{b,2}^2 K_\epsilon^{-1} + \lambda_{\rho,2}^2 K_\gamma^{-1} + 2 \lambda_{b,2} \lambda_{\rho,2} K_{\gamma \epsilon}^{-1}\nonumber\\
& & \lambda_{\sigma, 1}^2 K_\sigma + \lambda_{\sigma, 2}^2 K^{-1}_\sigma)
\eea
In Table \ref{exponentlist_spinful}, we first give the scaling of 
 the single-particle operators $b$ and $f_{\ua/\da}$, as well as the 
  standard order parameters of a LL of spinful fermions, i.e.
  $f_s^\dagger f_s$ and $f_\ua^\dagger f_\da$ (the $2k_f$-modes of these 
 operators correspond to CDW and SDW ordering, respectively), the pairing
 operators  $f_\ua f_\da$ (singlet) and $f_s f_s$ (triplet), and
 the order parameter of the Wigner crystal, $f^\dagger_\ua f^\dagger_\da f_\da f_\ua$.
 Apart from these standard order parameters of LL theory, we consider 
 a wide class of operators of the form $(f_\ua^\dagger)^m f_\ua^n (f_\da^\dagger)^p f_\da^q (b^\dagger)^r b^s$.
\section{Polaron Effects}\label{A_polaron}
In this subsection we will verify that the conventional construction of polaron
operators based on the canonical polaron transformation (CPT) \cite{mott} 
is 
equivalent to the construction within the bosonization approach presented
in Section \ref{single_p} of this paper.
We will also discuss the behavior of the polarization parameter $\lambda_c$.
The  CPT operator is given by
\bea
U_\lambda & = & e^{-\lambda\sum_{{k}\neq 0}\left(
F_{k}\beta_{k} n_{f, k}^\dagger-{\rm h.c.}\right)}
\eea
where $\beta_{k}$ is the phonon annihilation operator and $n_{f, k}$ is the fermion density
operator. $F_{k}$ is given by $F_k=\frac{1}{2}\sqrt{\frac{2\pi}{K_b|k|L}}$, and $\lambda$
specifies the strength of the phonon dressing. 
\begin{widetext}
\begin{table}
\begin{tabular}{|c|l|l|c|}  
\hline
operator $O(x)$ & wavevector $q$ & exponent $\tilde{\alpha}$ & remarks\\
\hline
\hline
\hline
$b$ & $2m k_b$ & $\frac{1}{2}\left[(2m)^2K_\delta+ K_\epsilon^{-1}\right]$ & \\ 
\hline
$f$ & $ (2n+1)k_f$ & $\frac{1}{2}[(2 n+1)^2K_\beta +K^{-1}_\gamma]$   &  \\
\hline
$f^\dagger f$ & $2nk_f$ & $2 n^2K_\beta$   & $n\neq 0$ \\
\hline
$f f$ & $2nk_f$ & $2 n^2K_\beta +2K^{-1}_\gamma$   & \\
\hline
$b^\dagger b$ & $2mk_b$ & $2 m^2 K_\delta$ & $m\neq 0$ \\
\hline
$[b^\dagger]^{m_1}$ & $2m k_b$ & $\frac{1}{2}\left[(2m)^2K_\delta+m_1^2 K_\epsilon^{-1}\right]$ & \\ 
\hline
$[f]^{2n_1}$ & $2n k_f$ & $\frac{1}{2}\left[(2n)^2K_\beta+(2n_1)^2K_\gamma^{-1}\right]$ & \\
\hline
$b^\dagger b f^\dagger f$ & $2mk_b+2nk_f$ &  $\frac{1}{2}\left[(2m)^2 K_\delta+(2n)^2K_\beta-
2(2m)(2n)K_{\beta\delta}\right]$ & $m,n\neq 0$ \\
\hline
$\left[ b^\dagger \right]^{m_1} f^\dagger f$ & $2mk_b+2nk_f$ & 
$\frac{1}{2}\left[(2m)^2 K_\delta+(2n)^2K_\beta-
2(2m)(2n)K_{\beta\delta}+m_1^2 K_\epsilon^{-1}\right]$ & $n\neq 0$\\
\hline
$\left[ b^\dagger \right]^{m_1} \left[ f\right]^{2n_1}$ & $2mk_b+2nk_f$ & 
 $\frac{1}{2}\left[(2m)^2 K_\delta+(2n)^2K_\beta-
2(2m)(2n)K_{\beta\delta}\right.$ & \\
 & & $\ \ \left.+m_1^2 K_\epsilon^{-1}+(2n_1)^2K_\gamma^{-1}
-2m_1(2n_1)K_{\gamma\epsilon}^{-1}\right]$ & \\
\hline
$\left[ b^\dagger \right]^{m_1} \left[ f\right]^{2n_1+1}$ & 
$2mk_b+(2n+1)k_f$ &  $\frac{1}{2}\left[(2m)^2 K_\delta+(2n+1)^2K_\beta-
2(2m)(2n+1)K_{\beta\delta}\right.$ & \\
 & & $\ \ \left.+m_1^2 K_\epsilon^{-1}+(2n_1+1)^2K_\gamma^{-1}
-2m_1(2n_1+1)K_{\gamma\epsilon}^{-1}\right]$ & \\
\hline
$\tilde{f}$ & $ (2n+1)k_f$ & $2- \frac{1}{2}[(2 n+1)^2K_\beta +K^{-1}_\beta]$   &  \\
\hline
$\tilde{f} \tilde{f}$ & $2nk_f$ & $2 n^2K_\beta +2K^{-1}_\beta$   & \\
\hline
$\tilde{b}$ & $2m k_b$ & $2-\frac{1}{2}\left[(2m)^2K_\delta+ K_\delta^{-1}\right]$ & \\ 
\hline
\end{tabular}
\caption{Table of Luttinger exponents and characteristic wavevectors
of various operators composed of boson and fermion operators.
$m$ and $n$ are arbitrary integer number resulting from the
higher order harmonics of the bosonization representations. 
%$O_{CDW}$ corresponds to the $2 k_f$-mode of $f^\dagger f$, $O_{f-PP}$ corresponds to the $(m=0, n=0)$-mode of $(b^\dagger)^{m_1} f^2$ with $m_1=\lambda_c$. 
}
\label{exponent_list}
\end{table}
\end{widetext}
The sum over the wavevector $k$ is over the regime 
 of acoustic modes.
When applied to a
fermion operator, the CPT gives 
\bea 
U^{-1}_\lambda f(x) U^{}_\lambda & = & f(x) e^{-\lambda \sum_{{k}\neq 0} (F_k \beta_k e^{-i{k}\cdot{r}}-{\rm h.c.})}
\label{polaron1}
\eea
 which is the standard expression for a fermionic polaron operator \cite{mott}.
%
%
%
%
 %We will now establish the equivalence
 %of this expression to the definition
 %$\tilde{f}(x) = f(x) e^{-i\lambda_c \Phi_b}$,
 %that we use in this paper.
In terms of Luttinger bosons the fermion density $n_{f, k}$ is given by:
\bea
n_{f, k} & = & \sqrt{{k L}/{2 \pi}} (B_k + B_{-k}^{\dagger}),
\eea
 and therefore 
%With this expression 
$U_\lambda$ becomes:
\bea
U_\lambda & = & e^{-\frac{\lambda}{2 \sqrt{K_b}} \sum_{{k}\neq 0} \left(B_k + B_{-k}^{\dagger}\right) \left(\beta_{-k} - \beta_k^{\dagger}\right)}
\label{U_lambda_bos}
\eea
When applied to a fermion operator, this transformation gives
\bea
U^{-1}_\lambda f(x) U_\lambda & = & f(x) e^{-i\lambda\Phi_b}
\label{polaron2}
\eea
because $U^{-1}_\lambda \Phi_f U_\lambda = \Phi_f - \lambda \Phi_b$,
 which can be seen from the operator representation (\ref{Phi}) and (\ref{Phi_F}), and by applying
the Baker-Hausdorff formula.
 The superficial difference between Eq. (\ref{polaron1}) and Eq. (\ref{polaron2}) can be overcome by a trivial phaseshift $\beta_k \rightarrow i {\rm sgn}(k) \beta_k$.
As discussed in this paper, $\lambda_c=K_{\epsilon}/K_{\gamma\epsilon}$ is
 the most appropiate choice for 
  the fermionic polarons, because the correlation 
 function of such polarons has the slowest algebraic decay. 
%$\lambda = \lambda_c$.
%
%
%
%
\begin{figure}[b]
\includegraphics[width=4cm]{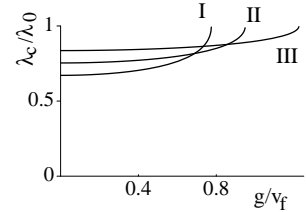}
\caption{Plot of the 'dressing' parameter $\lambda_c$ compared to complete dressing $\lambda_0$, for $K_b =5$ and $G/v_f =0.1$ as a function of $g/v_f$. 
With $v_b/v_f=2,3,5$ for $I$--$III$.  }
\label{dressing}
\end{figure}
%The connection between the polaron transformation and the exact diagonalization is given by
%\bea
%U_{diag} & = & U_{\alpha} U_{\lambda_c}
%\eea
%where $U_\alpha$ is a pure rotation of the form
%\bea
%\Pi_F  =  \cos \alpha \Pi_A + \sin \alpha \Pi_a, & &
%\Phi_F  =  \cos \alpha \Phi_A + \sin \alpha \Phi_a, \nonumber\\
%\Pi_B  =  - \sin \alpha \Pi_A + \cos \alpha \Pi_a, & &
%\Phi_B  =  - \sin \alpha \Phi_A + \cos \alpha \Phi_a \nonumber
%\eea
%Note that a rotation does not change the exponents in the correlation functions, so the resulting Luttinger parameters are entirely due to the polaron transformation. In an analogous way, one can use a b-polaron transformation and a rotation to obtain the same result. 
In Fig. \ref{dressing} we plot this polarization
 parameter $\lambda_c$ in comparison to the value of 'complete dressing' $\lambda_0$:
\bea
\lambda_0 & = & \frac{2 g}{v_b} \sqrt{K_b}
\label{lambda_0}
\eea
To obtain this quantity, which
 is the 'size' the polarization would have if the fermions were static, we set the Fermi velocity to zero ($v_f = 0$), so that the Hamiltonian of the system becomes:
\bea
H = \sum_{k \neq 0} v_b |k| \beta_k^{\dagger} \beta_k + \sum_{k \neq 0} g |k| (\beta_k ^{\dagger} + \beta_{-k}) (B_k + B_{-k} ^{\dagger})
\label{Ham_static}
\eea
In this limit of infinitely heavy fermions, 
 a term of the type (\ref{eq:H_f^2}) does not exist.
In (\ref{Ham_static}) the linear term simply 
shifts the bosonic modes by an amount $g/v_b (B_k + B_{-k}^{\dagger})$.
As we can see from (\ref{U_lambda_bos}) this corresponds 
exactly to the polarization $\lambda_0$, Eq. (\ref{lambda_0}).
However, for finite hopping, or finite mass of the fermions, 
 this value is reduced, because the polarization cloud cannot entirely 'follow' the fermionic atoms. For $U_{bf}\rightarrow 0$ one finds
\bea
\lambda_c/\lambda_0 & \rightarrow & \frac{v_b}{v_b + v_f}
\label{lambda_limit}
\eea
In Fig. \ref{dressing} we show $\lambda_c$ plotted  for different sets 
of parameters. It smoothly interpolates between the small 
interaction limit (\ref{lambda_limit}) and the static limit (\ref{lambda_0}).
 The static limit is achieved in the vicinity of phase separation
 because the fermionic effective mode velocity
 $v_A$ goes to zero when the system approaches separation. This corresponds to 
 infinitely heavy fermions, and therefore to the static limit.

%
%
%
%%
%
%We want to point out that
%although the physical picture of a polaron is very intuitive, it is in general a non-trivial task to determine the actual 'size' of the screening cloud. In the model that we discussed in this paper this has been done analytically by finding $\lambda_c$.

%
\begin{widetext}
\begin{table}
\begin{tabular}{|c|l|l|c|}  
\hline
operator $O(x)$ & wavevector $q$ & exponent $\tilde{\alpha}$ & remarks\\
\hline
\hline
$b$ & $2mk_b$ & $\frac{1}{2}[(2m)^2K_\delta +K_\epsilon^{-1}]$  &  \\
\hline
$f_{\ua/\da}$ & $(2n+1)k_f$ & $(n+\frac{1}{2})^2(K_\beta+K_\sigma) +\frac{1}{4}(K_\gamma^{-1}+K_\sigma^{-1})$  & \\
\hline
$f^\dagger_s f_s$ & $2nk_f$ & $\left[n^2 K_\beta+n^2K_\sigma\right]$ & $n \neq 0$\\
\hline
$f^\dagger_\ua f_\da$ & $2(n+m)k_f$ & $\left[(n+m)^2 K_\beta+(n-m-1)^2 K_\sigma + K_\sigma^{-1}\right]$ & $n \neq 0$\\
\hline
$f_s f_s$ & $2nk_f$ & $\left[n^2 K_\beta+n^2K_\sigma + K^{-1}_\gamma + K^{-1}_\sigma \right]$ &\\
\hline
$f_\ua f_\da$ & $2(n+l+1)k_f$ & $\left[(n+l+1)^2 K_\beta+K^{-1}_\gamma + (n-l)^2K_\sigma\right]$ &\\
\hline
$f_\ua^\dagger f_\da^\dagger f_\da f_\ua$ & $2(n+m)k_f$ & $\left[(n+m)^2 K_\beta + (n-m)^2K_\sigma\right]$ &\\
\hline
$b^\dagger b$ & $2mk_b$ & $2m^2K_\delta$  & $m\neq 0$ \\
\hline
$[b^\dagger]^{m_1}$ & $2m k_b$ &  $\left[2 m^2K_\delta+\frac{1}{2}m_1^2 K_\epsilon^{-1}\right]$ & \\ 
 \hline
$[f_s]^{2n_1}$ & $2n k_f$ &  $\left[n^2(K_\beta+K_\sigma)+n_1^2 (K_\gamma^{-1}
+K_{\sigma}^{-1})\right]$ & \\
\hline
$[f_s]^{2n_1+1}$ & $(2n+1) k_f$ & $\left[(n+\frac{1}{2})^2 (K_\beta+K_{\sigma})
+(n_1+\frac{1}{2})^2(K_\gamma^{-1}
+K_{\sigma}^{-1})\right]$ & \\
\hline
%$b^\dagger b f^\dagger f$ & $2mk_b+2nk_f$ & $2-\frac{1}{2}\left[
%(2m\delta_1+2n\beta_1)^2+(2m\delta_2+2n\beta_2)^2\right]$ & $m,n\neq 0$ \\
% & & $=2-\frac{1}{2}\left[(2m)^2 K_\delta+(2n)^2K_\beta-
%2(2m)(2n)K_{\beta\delta}\right]$ & \\
%\hline
%$b^\dagger b \left[ f \right]^{2n_1}$ & $2mk_b+2nk_f$ & $2-\frac{1}{2}\left[
%(2m\delta_1+2n\beta_1)^2+(2m\delta_2+2n\beta_2)^2
%+(2n_1)^2(\gamma_1^2+\gamma_2^2)\right]$ & $m\neq 0$ \\
% & & $=2-\frac{1}{2}\left[(2m)^2 K_\delta+(2n)^2K_\beta-
%2(2m)(2n)K_{\beta\delta}+(2n_1)^2 K_\gamma\right]$ & \\
%\hline
%$b^\dagger b \left[ f \right]^{2n_1+1}$ & $2mk_b+(2n+1)k_f$ & 
%$2-\frac{1}{2}\left[
%(2m\delta_1+(2n+1)\beta_1)^2+(2m\delta_2+(2n+1)\beta_2)^2
%+(2n_1+1)^2(\gamma_1^2+\gamma_2^2)\right]$ & $m\neq 0$ \\
% & & $=2-\frac{1}{2}\left[(2m)^2 K_\delta+(2n+1)^2K_\beta-
%2(2m)(2n+1)K_{\beta\delta}+(2n_1+1)^2 K_\gamma\right]$ & \\
%\hline
%\hline
%
$\left[ b^\dagger \right]^{m_1} \left[ f_s\right]^{2n_1}$ & $2mk_b+2nk_f$ & 
$\left[2m^2 K_{\delta}+n^2K_{\beta}-
2\sqrt{2} m n K_{\beta \delta}+n^2 K_{\sigma}\right.$ & \\
 & & $\ \ \left.+\frac{1}{2}m_1^2 K_{\epsilon}^{-1}+n_1^2K_{\gamma}^{-1}
-\sqrt{2}m_1 n_1 K_{\gamma \epsilon}^{-1}+n_1^2 K_{\sigma}^{-1}\right]$ & \\
\hline
$\left[ b^\dagger \right]^{m_1} \left[ f\right]^{2n_1+1}$ & 
$2mk_b+(2n+1)k_f$ & 
$\left[2m^2 K_{\delta}+(n+\frac{1}{2})^2K_{\beta}-
\sqrt{2} m (2n+1) K_{\beta \delta}+(n+\frac{1}{2})^2 
K_{\sigma}\right.$ & \\
 & & $\ \ \left.+\frac{1}{2}m_1^2 K_{\epsilon}^{-1}
+(n_1+\frac{1}{2})^2K_{\gamma}^{-1}
-\sqrt{2}m_1 (n_1+\frac{1}{2}) K_{\gamma \epsilon}^{-1}
+(n_1+\frac{1}{2})^2 K_{\sigma}^{-1}\right]$ & \\
\hline
$\left[ b^\dagger \right]^{m_1} 
\left[ f_\uparrow f_\downarrow\right]$ & $2mk_b+2(n+l+1)k_f$ & 
$\left[2m^2 K_{\delta}+(n+l+1)^2K_{\beta}-
2\sqrt{2} m (n+l+1) K_{\beta\delta}+(n-l)^2 K_{\sigma}\right.$ & \\
 & & $\ \ \left.+\frac{1}{2}m_1^2 K_{\epsilon}^{-1}+K_{\gamma}^{-1}
-\sqrt{2}m_1 K_{\gamma \epsilon}^{-1}\right]$ & \\
\hline
$\tilde{b}$ & $2mk_b$ &  $\frac{1}{2}[(2m)^2K_\delta +K_\delta^{-1}]$   &  \\
\hline
$\tilde{f}_{\ua/\da}$ & $(2n+1)k_f$ & $(n+\frac{1}{2})^2(K_\beta+K_\sigma)+ \frac{1}{4}(K_\beta^{-1}+K_\sigma^{-1})$  & \\
\hline
$\tilde{f}_s \tilde{f}_s$ & $2nk_f$ & $\left[n^2 K_\beta+n^2K_\sigma + K^{-1}_\beta + K^{-1}_\sigma \right]$ &\\
\hline
$\tilde{f}_\ua \tilde{f}_\da$ & $2(n+l+1)k_f$ & $\left[(n+l+1)^2 K_\beta+K^{-1}_\beta + (n-l)^2K_\sigma\right]$ &\\
\hline
\end{tabular}
\caption{Table of Luttinger exponents and characteristic wavevectors
of various order parameters composed by boson and fermion operators
for spinful system.
$m$, $n$ and $l$ are arbitrary integer number resulting from the
higher order harmonics of the bosonization representations.}
\label{exponentlist_spinful}
\end{table}
\end{widetext}
%
%
%
%
%%%%%%%%%%%%%%%%%%%%%%%%%
\section{RG flow of an impurity term}
\label{impurity}

We consider the following local impurity  term  in our Hamiltonian:
%Before concluding our discussion of BFM with spinless fermions we
%discuss an approach for observing the quantum phase transition between
%the CDW and the $f$-PP phases in laser stirring experiments \cite{stirring}.
%When a laser beam is focused at the center of the BFM cloud, it can
%create a local potential for fermionic atoms.  In the $f$-PP phase the
%stirring potential can be moved through the system with no
%dissipation, provided that its motion is slower than some critical
%velocity (experimental techniques are
%discussed in Ref. \onlinecite{stirring}). At the $f$-PP/CDW phase
%boundary the critical velocity goes to zero, reflecting a transition
%to the insulating (CDW) state.  This scenario follows from considering
%the effect of a small local pinning potential for fermions.
%
%
%
%
%%
\bea
H_{imp} & = & \int dx V(x) f^{\dagger}(x) f(x)
\eea
with $V(x)$ strongly spatially peaked around at $x=0$.
%, which we set to zero,
% $x_0 = 0$.
This term leads to an additional term in the action given by
\bea
S_{imp} & = & \sum_m \frac{v_m}{2} \int d\tau \exp(2 i m \Theta_f)
\eea
where $v_m$ is the Fourier transform of the potential 
 $V(x)$ around $2 m k_f$. The RG flow 
of these terms at one-loop, as discussed in
 \cite{kane} and \cite{gnt} is given by:
\bea
\frac{d v_m}{d l} & = & (1 - m^2 K_{\beta}) v_m
\eea
Therefore, such an
impurity term 
  is relevant exactly for $K_{\beta} < 1$, and irrelevant outside of the CDW phase.
The physical interpretation of this result is as follows:
 In the CDW regime a local impurity 'pins' 
 the charge ordering of the system. 
%The very definition of CDW QLRO
% is the divergence of the CDW susceptibility. Therefore 
% any perturbation that couples to that channel -- like the impurity term here --
% will generate ordering. 
%
%However, in the hypothetical metallic phase this term should be relevant, since
Outside of the CDW phase, 
 the irrelevance of an impurity term in the system is 
an indication of a superfluid phase, which 
 must be provided by polaron pairing.

\end{document}